\documentclass{gji}
\usepackage{timet}
\usepackage[dvips]{graphicx}
\usepackage{natbib}

\title[Free and smooth boundaries for elastic waves]
  {Free and smooth boundaries in 2-D finite-difference schemes for transient elastic waves}

\author[B. Lombard, J. Piraux, C. G\'elis, J. Virieux]
  {B. Lombard$^1$, J. Piraux$^1$, C. G\'elis$^2$, J. Virieux$^2$ \\
  $^1$ CNRS, Laboratoire de M\'ecanique et d'Acoustique, Marseille, France\\
  $^2$ CNRS, G\'eosciences Azur, Sophia-Antipolis, France.\\
  }

\date{Received 200?; in original form 200?}
\pagerange{\pageref{firstpage}--\pageref{lastpage}}
\volume{?}
\pubyear{200?}

\begin{document}

\label{firstpage}

\maketitle

%-----------------------------------

\begin{summary}
A method is proposed for accurately describing arbitrary-shaped free boundaries in finite-difference schemes for elastodynamics, in a time-domain velocity-stress framework. The basic idea is as follows: fictitious values of the solution are built in vacuum, and injected into the numerical integration scheme near boundaries. The most original feature of this method is the way in which these fictitious values are calculated. They are based on boundary conditions and compatibility conditions satisfied by the successive spatial derivatives of the solution, up to a given order that depends on the spatial accuracy of the integration scheme adopted. Since the work is mostly done during the preprocessing step, the extra computational cost is negligible. Stress-free conditions can be designed at any arbitrary order without any numerical instability, as numerically checked. Using 10 grid nodes per minimal S-wavelength with a propagation distance of 50 wavelengths yields highly accurate results. With 5 grid nodes per minimal S-wavelength, the solution is less accurate but still acceptable. A subcell resolution of the boundary inside the Cartesian meshing is obtained, and the spurious diffractions induced by staircase descriptions of boundaries are avoided. Contrary to what occurs with the vacuum method, the quality of the numerical solution obtained with this method is almost independent of the angle between the free boundary and the Cartesian meshing. 
\end{summary}


%-----------------------------------

\section{Introduction}\label{SecIntro}

Various approaches have been proposed for simulating the propagation of elastic waves with free boundaries. The first approach is based on variational methods, as done in finite elements \citep{DAY77}, spectral finite elements \citep{KOMATITSCH98} and discontinuous Galerkin \citep{BENJEMAA07}. These methods provide a fine geometrical description of boundaries by adapting the mesh to the boundaries. Boundary conditions are accounted for weakly by the underlying variational formulation. However, a grid-generating tool is required, and small time steps may result from the smallest geometrical elements and from the stability condition. The SAT methods based on energy estimates \citep{CARPENTER94} avoid these limitations by introducing Cartesian grids and give time-stable high-order schemes with interfaces. However and up to our knowledge, these methods have not been applied so far to elastodynamics with free boundaries. 

The second approach used in this context is based on the strong form of elastodynamics, as done in finite differences and spectral methods \citep{TESSMER94}. In seismology, finite differences are usually implemented on staggered Cartesian grids, either with completely staggered stencils (CSS) or with the recently developed partially staggered stencils (PSS). With CSS, the velocity and stress components are distributed between different node positions \citep{VIRIEUX86}. With PSS, all the velocity components are computed at a single node, as are the stress components, although the latter are shifted by half a node in two separate grids. Second-order \citep{SAENGER00, SAENGER04} and fourth-order \citep{BOHLEN03,CRUZ04} spatially-accurate PSS have been developed; for further discussion, we denote them PSS-2 and PSS-4, respectively. Unlike variational methods, finite differences require special care to incorporate the free boundary conditions strongly. There exist two main strategies for this purpose:
\begin{enumerate}
\item First, the boundaries can be taken into account implicitly by adjusting the physical parameters  locally \citep{KELLY76,VIRIEUX86,MUIR92}. The best-known implicit approach is the so-called \textit{vacuum method} \citep{ZAHRADNIK95,GRAVES96,MOCZO02,GELIS05}. For instance, the vacuum method applied to PSS involves setting the elastic parameters in the vacuum to zero, and using a small density value in the first velocity node in the vacuum to avoid a division by zero. However, this easy-to-implement method gives at best second-order spatial accuracy. In addition, a systematic numerical study has shown that the accuracy of the solution decreases dramatically when the angle between the boundary and the meshing increases \citep{BOHLEN06}. Lastly, applying the vacuum method sometimes gives rise to instabilities: see for instance PSS-4 \citep{BOHLEN03}.  
\item A second idea is to explicitly change the scheme near the boundaries \citep{KELLY76}. The best-known explicit approach is the so-called \textit{image method}, which was developed for dealing with flat boundaries to fourth-order accuracy \citep{LEVANDER88} and then extended to variable topographies \citep{JIH88,ROBERTSSON96,ZHANGW06}. However image methods require a fine grid to reduce the spurious diffractions up to an acceptable level. To avoid this spatial oversampling, various techniques have been proposed, such as grid refinement in the vicinity of the boundary \citep{RODRIGUES93} or adjusted finite-difference approximations: see \citep{MOCZO07} for a review on these subjects.
\end{enumerate}

The aim of this paper is to present a finite-difference approach accounting for free boundaries without introducing the aforementioned drawbacks of the vacuum and image methods. The requirements are as follows: smooth arbitrary-shaped boundaries must be treated as simply as straight boundaries; the accuracy of the method must not depend on the position of the boundary relative to the meshing; and lastly, the computations must be stable even with very long integration times. We establish that these requirements can be met by applying an explicit approach involving fictitious values of the solution in the vacuum. In previous studies, interface problems in the context of  elastodynamics were investigated in a similar way \citep{PIRAUX01,LOMBARD04,LOMBARD06}. The fictitious values are high-order Taylor expansions of the boundary values of the solution. Estimating these boundary values involves some mathematical background, in order to compute the high-order boundary conditions; to determine a minimal set of independent boundary values; lastly, to perform a least-square numerical estimation of this minimal set. To help the reader, subroutines in FORTRAN are proposed freely at the web page {\tt http://w3lma.cnrs-mrs.fr/\textasciitilde MI/Software/}. These subroutines enable a straightforward implementation of the algorithms detailed in the present paper.

The disadvantage here is that the above requirements cannot be fully satisfied if staggered-grid schemes are used. Single-grid finite-difference schemes are therefore chosen, where all the unknowns are computed at the same grid nodes. Our numerical experiments are based on the high-order ADER schemes which are widely used in aeroacoustics \citep{MUNZ05}. Although these schemes are not yet widely used in the field of seismology \citep{DUMBSER06}, they have also great qualities because of their accuracy and their stability properties: using 10 grid nodes per minimal S-wavelength with a propagation distance of 50 wavelengths gives highly accurate results. Moreover, on Cartesian grids, these methods do not require much more computational memory than staggered-grid schemes.

This paper is organized as follows. Section 2 deals with the continuous problem: the high-order boundary conditions and compatibility conditions are stated. These conditions are useful for handling the discrete problem presented in section 3, where the focus is on obtaining fictitious values of the solution in the vacuum. In section 4, numerical experiments confirm the efficiency of this method in the case of various topographies. In section 5, conclusions are drawn and some prospects suggested.

\section{The continuous problem}\label{SecContinuous}

\subsection{Framework}\label{SubSecFrame}

\begin{figure}
\begin{center}
\includegraphics[scale=0.8]{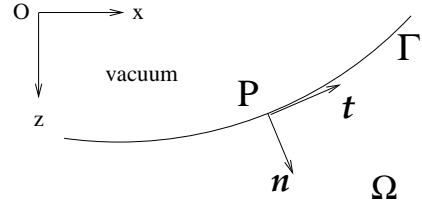}
\caption{{\it Boundary $\Gamma$ between a solid $\Omega$ and vacuum}.}
\label{FigPatate}
\end{center}
\end{figure}

Let us consider a solid $\Omega$ separated from the vacuum by a boundary $\Gamma$ (Figure \ref{FigPatate}). The configuration is in-plane and two-dimensional, with a horizontal axis $x$ and a vertical axis $z$ pointing respectively rightward and downward. The boundary $\Gamma$ is described by a parametric expression $(x(\tau),\,z(\tau))$ where the parameter $\tau$ describes the sampling of the boundary. The tangential vector and the normal vector are $\mitbf{t}=\,^T(x^{'}(\tau),\,z^{'}(\tau))$ and $\mitbf{n}=\,^T(-z^{'}(\tau),\,x^{'}(\tau))$, respectively, with $x^{'}(\tau)=\frac{d\,x}{d\,\tau}(\tau)$, $z^{'}(\tau)=\frac{d\,z}{d\,\tau}(\tau)$, and $T$ refers to the transposed vector. We assume the spatial derivatives at any point of the boundary to be available, as specified below. 

The solid $\Omega$ is assumed to be linearly elastic, isotropic, and to have the following constant physical parameters: the density $\rho$ and the Lam\'e coefficients $\lambda$, $\mu$. The P- and S-wave velocities are $c_p=\sqrt{(\lambda+2\,\mu)/\rho}$ and $c_s=\sqrt{\mu/\rho}$. A velocity-stress formulation is adopted, hence the unknowns are the horizontal and vertical components of the elastic velocity $\mitbf{v}=\,^T(v_x,\,v_z)$, and the independent components of the elastic stress tensor $\mitbf{\sigma}=\,^T(\sigma_{xx},\,\sigma_{xz},\,\sigma_{zz})$. Setting
$$
\begin{array}{l}
\displaystyle
\mitbf{A}=
\left(
\begin{array}{ccccc}
0 & 0 & \hspace{0.1cm}1/\rho & 0 & 0\\
0 & 0 & 0 & 1/\rho & 0\\
\lambda+2\,\mu & 0 & 0 & 0 & 0\\
0 & \mu & 0 & 0 & 0 \\
\lambda & 0 & 0 & 0 & 0
\end{array}
\right),\\
\\
\displaystyle
\mitbf{B}=
\left(
\begin{array}{ccccc}
0 & 0 & 0 & \hspace{0.1cm}1/\rho & 0\\
0 & 0 & 0 & 0 & 1/\rho\\
0 & \lambda & 0 & 0 & 0\\
\mu & 0 & 0 & 0 & 0\\
0 & \lambda+2\mu & 0 & 0 & 0
\end{array}
\right), 
\end{array}
$$ 
the solution $\mitbf{U}=\,^T(v_x,\,v_z,\,\sigma_{xx},\,\sigma_{xz},\,\sigma_{zz})$ satisfies the first-order hyperbolic system \citep{VIRIEUX86}
\begin{equation}
\frac{\textstyle \partial}{\textstyle \partial\,t}\,\mitbf{U}=\mitbf{A}\,\frac{\textstyle \partial}{\textstyle \partial\,x}\,\mitbf{U}+\mitbf{B}\,\frac{\textstyle \partial}{\textstyle \partial\,z}\,\mitbf{U}.
\label{LC}
\end{equation}

\subsection{High-order boundary conditions}\label{SubSecBC}

At any point $P(\tau)$ on the free surface $\Gamma$ (Figure \ref{FigPatate}), the stress tensor satisfies the homogeneous Dirichlet conditions $\mitbf{\sigma}.\mitbf{n}=\mitbf{0}$. These zero-th order boundary conditions are written compactly
\begin{equation}
\mitbf{L}^0(\tau)\,\mitbf{U}^0(x(\tau),\,z(\tau),\,t)=\mitbf{0},
\label{L0U0}
\end{equation}
where $\mitbf{U}^0$ is the limit value of $\mitbf{U}$ at $P$ and $\mitbf{L}^0$ is the matrix
$$
\mitbf{L}^0(\tau)=
\left(
\begin{array}{ccccc}
0 & 0 & -z^{'}(\tau) & x^{'}(\tau) & 0 \\
0 & 0 & 0 & -z^{'}(\tau) & x^{'}(\tau) 
\end{array}
\right).
$$
From now on, the dependence on $\tau$ is generally omitted. To determine the boundary conditions satisfied by the first-order spatial derivatives of $\mitbf{U}$, two tasks are performed. First, the zeroth-order boundary conditions (\ref{L0U0}) are differentiated in terms of $t$. The time derivative is replaced by spatial derivatives using the conservation laws (\ref{LC}), which gives
\begin{equation}
\mitbf{L}^0\left(\mitbf{A}\,\frac{\textstyle \partial}{\textstyle \partial\,x}\,\mitbf{U}^0+\mitbf{B}\,\frac{\textstyle \partial}{\textstyle \partial\,z}\,\mitbf{U}^0\right)=\mitbf{0}.
\label{L0t}
\end{equation}
Secondly, the zeroth-order boundary conditions (\ref{L0U0}) are differentiated in terms of the parameter $\tau$ describing $\Gamma$. The chain-rule gives
\begin{equation}
\left(\frac{\textstyle d}{\textstyle d\,\tau}\,\mitbf{L}^0\right)\,\mitbf{U}^0+\mitbf{L}^0\left(x^{'}\frac{\textstyle \partial}{\textstyle \partial\,x}\,\mitbf{U}^0+z^{'}\frac{\textstyle \partial}{\textstyle \partial\,z}\,\mitbf{U}^0\right)=\mitbf{0}.
\label{L0tau}
\end{equation}
Since the matrix $d\,\mitbf{L}^0/d\,\tau$ in (\ref{L0tau}) involves $x^{''}$ and $z^{''}$, it accounts for the curvature of $\Gamma$ at $P$. Setting the block matrix
$$
\mitbf{L}^1=
\left(
\begin{array}{ccc}
\mitbf{L}^0  &  \mitbf{0}  &  \mitbf{0}\\
[3pt]
\mitbf{0}    &  \mitbf{L}^0\,\mitbf{A}  &  \mitbf{L}^0\,\mitbf{B}  \\
\displaystyle
\frac{\textstyle d}{\textstyle d\,\tau}\,\mitbf{L}^0 & x^{'}\mitbf{L}^0 & z^{'}\mitbf{L}^0
\end{array}
\right),
$$
equations (\ref{L0U0}), (\ref{L0t}) and (\ref{L0tau}) give the boundary conditions up to the first-order 
$$
\mitbf{L}^1\,\mitbf{U}^1=\mitbf{0},
$$
with 
$$
\mitbf{U}^1=\lim_{M\in \Omega \rightarrow P}
\,^T\left(
^T\mitbf{U},
\frac{\textstyle \partial}{\textstyle \partial\,x}\,^T\mitbf{U},\,\frac{\textstyle \partial}{\textstyle \partial\,z}\,^T\mitbf{U}\right).
$$
Let $k\geq 1$ be an integer whose value will be discussed in section \ref{SecDiscrete}. To get the boundary conditions up to the $k$-th order, one deduces from (\ref{L0U0}) 
\begin{equation}
\frac{\textstyle \partial^k}{\textstyle \partial\,\tau^{k-\alpha}\,\partial\,t^\alpha}\,\mitbf{L}^0\,\mitbf{U}^0=\mitbf{0},\qquad \alpha=0,...,k.
\label{DTauDt}
\end{equation}
The $\tau$-derivatives are replaced by spatial derivatives by applying $(k-\alpha)$-times the chain rule. The $t$-derivatives are replaced by spatial derivatives by injecting $\alpha$-times the conservation laws (\ref{LC}). The boundary conditions so-obtained up to the $k$-th order can be written compactly
\begin{equation}
\mitbf{L}^k\,\mitbf{U}^k=\mitbf{0},
\label{LkUk}
\end{equation}
with
\begin{equation}
\displaystyle
\mitbf{U}^k=\lim_{M\in \Omega \rightarrow P}
\,^T\left(
^T\mitbf{U},
...,\,
\frac{\textstyle \partial^\alpha}{\textstyle \partial\, x^{\alpha-\beta}\,\partial\,z^\beta}\,^T\mitbf{U},
...,\,
\frac{\textstyle \partial^k}{\textstyle \partial\,z^k}\,^T\mitbf{U}
\right),
\label{Uk}
\end{equation}
where $\alpha=0,...,\,k$ and $\beta=0,...,\,\alpha$. The vector $\mitbf{U}^k$ has $n_v=5(k+1)\,(k+2)/2$ components. $\mitbf{L}^k$ is a $n_l \times n_v$ matrix, with $n_l=(k+1)\,(k+2)$. This matrix  involves the successive derivatives of the curvature of $\Gamma$ at $P$. Computing $\mitbf{L}^k$ with $k>2$ is a tedious task, which can be greatly simplified by using computer algebra tools. 

\subsection{Compatibility conditions}\label{SubSecCC}

The second spatial derivatives of stress components are linked together by the compatibility condition of Barr\'e-de Saint Venant \citep{LOVE}
\begin{equation}
\begin{array}{l}
\displaystyle
\frac{\textstyle \partial^2 \,\sigma_{xz}}{\textstyle \partial \,x\,\partial\,z}
=\alpha_2\,\frac{\textstyle \partial^2 \,\sigma_{xx}}{\textstyle \partial \,x^2}
+\alpha_1\,\frac{\textstyle \partial^2 \,\sigma_{zz}}{\textstyle \partial \,x^2}
+\alpha_1\,\frac{\textstyle \partial^2 \,\sigma_{xx}}{\textstyle \partial \,z^2}
+\alpha_2\,\frac{\textstyle \partial^2 \,\sigma_{zz}}{\textstyle \partial \,z^2},
\end{array}
\label{Barre}
\end{equation}
with
$$
\alpha_1=\frac{\textstyle \lambda+2\,\mu}{\textstyle 4\,(\lambda+\mu)},\qquad 
\alpha_2=-\frac{\textstyle \lambda}{\textstyle 4\,(\lambda+\mu)}.
$$
This compatibility condition is a necessary and sufficient condition for the strain tensor to be symmetrical. If $k\geq 2$, it can be differentiated $(k-2)$-times in terms of $x$ and $z$. With $k\geq 2$, one obtains $n_c=k\,(k-1)/2$ relations; with $k<2$, $n_c=0$. Unlike the boundary conditions, these compatibility conditions are satisfied everywhere in $\Omega$: in particular, they are satisfied at $P$ on $\Gamma$. The vector of boundary values $\mitbf{U}^k$ can therefore be expressed in terms of a shorter vector $\hat{\mitbf{U}}^k$ with $n_v-n_c$ independent components
\begin{equation}
\mitbf{U}^k=\mitbf{G}^k\,\hat{\mitbf{U}}^k. 
\label{MatG}
\end{equation}
An algorithm for building the $n_v\times (n_v-n_c)$ matrix $\mitbf{G}^k$ is given in \citet{LOMBARD06}. 

\section{The discrete problem}\label{SecDiscrete}

\subsection{Numerical scheme}\label{SubSecScheme}

To integrate the hyperbolic system (\ref{LC}), we introduce a single Cartesian lattice of grid points: $(x_i,z_j,t_n)=(i\,h,j\,h,\,n\,\Delta\,t)$, where $h$ is the mesh spacing and $\Delta\,t$ is the time step. Unlike with staggered grids, all the unknowns are computed at the same grid nodes. The approximation $\mitbf{U}_{i,j}^n$ of $\mitbf{U}(x_i,z_j,t_n)$ is computed using any explicit, two-step, and spatially-centred finite-difference scheme. A review of the huge body of literature on finite-differences is given in \citet{LEV90} and \citet{MOCZO07}. 

Here we propose to use ADER schemes, that allow to reach easily arbitrary high-order of time and space accuracy \citep{MUNZ05}. On Cartesian grids, these finite-volume integration schemes originally developed for aeroacoustic applications are equivalent to finite-difference Lax-Wendroff-type integration schemes \citep{LORCHER06}. In the numerical experiments described in section \ref{SecNum}, we use a fourth-order ADER integration scheme. This scheme is stable under the Courant-Friedrichs-Lewy (CFL) condition $c_p\,\Delta\,t/h \leq 0.9$ in 2D; as usually with single-grid schemes, it is slightly dissipative \citep{MUNZ05}. 

Many other single-grid schemes can be used in this context. In particular, the method described in the next subsections has been successfully combined with flux-limiter schemes \citep{LEV90} and with the standard second-order Lax-Wendroff scheme. Difficulties have been encountered with dissipative-free schemes based on centred staggered-grid finite-difference schemes, as we will see in section \ref{SubSecStag}.

\subsection{Use of fictitious values}\label{SubSecExtra1}

\begin{figure}
\begin{center}
\includegraphics[scale=0.5]{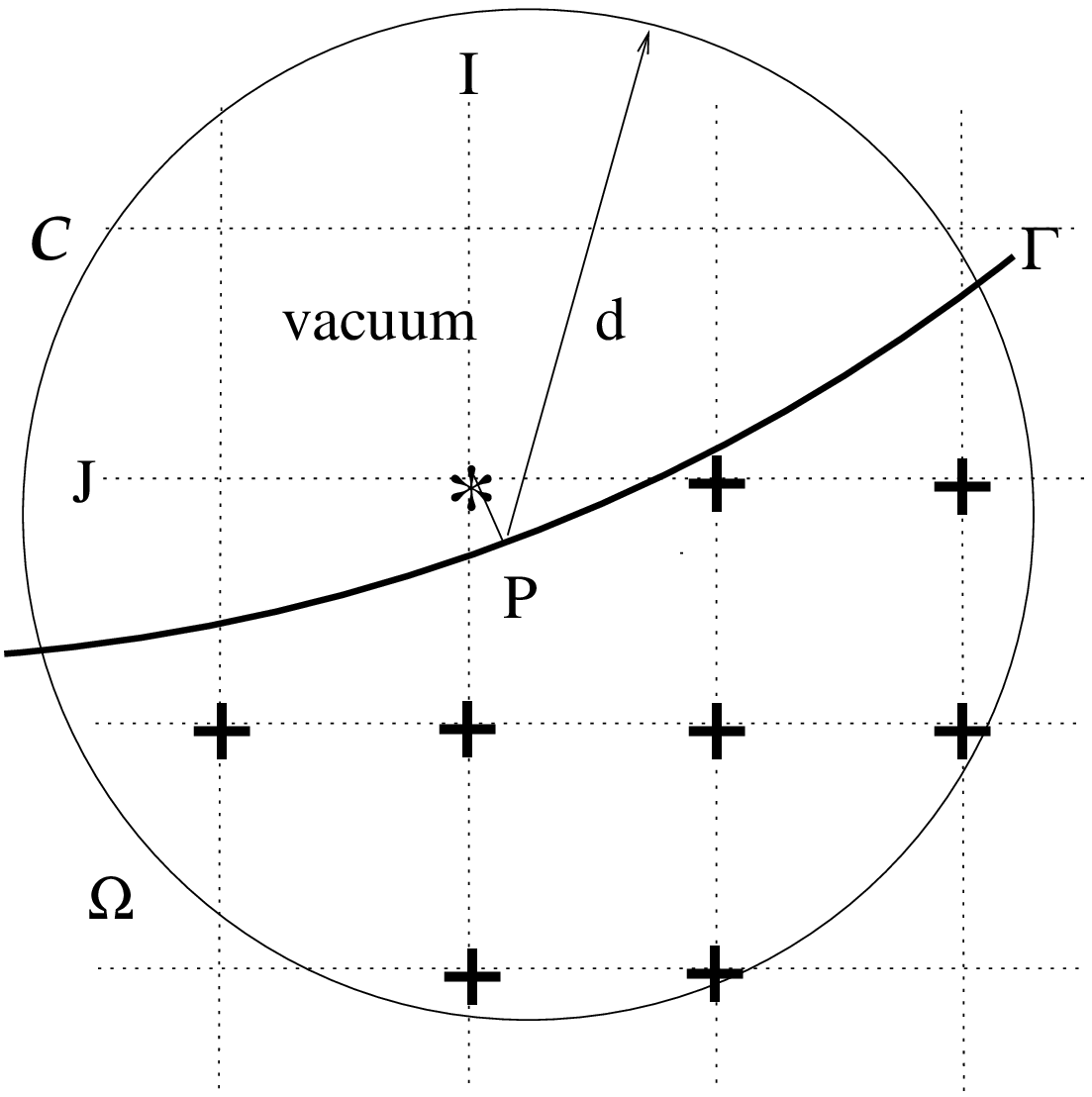}
\caption{\textit{Determination of the fictitious value $\mitbf{U}_{IJ}^*$ required for time-marching at neighboring grid nodes. P is the orthogonal projection of $(x_I,\,z_J)$ on $\Gamma$. The $n_p$ grid nodes in $\Omega$ and inside the circle ${\cal C}$ centred at $P$ with a radius $d$ are denoted by {\bf +}}.}
\label{FigSolMod}
\end{center}
\end{figure} 

Time-marching at grid-points where the stencil crosses $\Gamma$ requires fictitious values of the solution in the vacuum, which have to be determined. The question arises as to how to compute, for instance, the fictitious value $\mitbf{U}_{I,J}^*$ at the grid point $(x_I,\,z_J)$ in the vacuum, as sketched in Figure \ref{FigSolMod}. Let $P(\tau)$ be the orthogonal projection of $(x_I,\,z_J)$ on $\Gamma$, with coordinates $(x_P=x(\tau),\,z_P=z(\tau))$. At any grid point $(x_i,\,z_j)$, we denote
$$
\mitbf{\Pi}_{i,j}^k=\left(\mitbf{I}_5,
...,
\frac{\textstyle (x_i-x_P)^{\alpha-\beta}(z_j-z_P)^\beta}{\textstyle (\alpha-\beta)\,!\,\beta\,!}\mitbf{I}_5,
...,
\frac{\textstyle (z_j-z_P)^k}{\textstyle k\,!}\mitbf{I}_5\right)
$$
the $5\times n_v$ matrix containing the coefficients of $k$-th order Taylor expansions in space at $P$, where $\mitbf{I}_5$ is the $5\times5$ identity matrix, $\alpha=0,...,\,k$, and $\beta=0,...,\,\alpha$. The fictitious value $\mitbf{U}_{I,J}^*$ is defined as the Taylor-like extrapolation 
\begin{equation}
\mitbf{U}_{I,J}^*=\mitbf{\Pi}_{I,J}^k\,\mitbf{U}^k,
\label{SolMod0} 
\end{equation}
where $\mitbf{U}^k$ defined by (\ref{Uk}) still remains to be estimated. 

\subsection{Reduced vector of boundary values}\label{SubSecSVD}

Before determining $\mitbf{U}^k$ in (\ref{SolMod0}), we first reduce the number of independent components it contains. The expressions obtained in section \ref{SecContinuous} are used for this purpose. The linear homogeneous system following from (\ref{LkUk}) and (\ref{MatG}) is
\begin{equation}
\mitbf{L}^k\,\mitbf{G}^k\,\hat{\mitbf{U}}^k=\mitbf{0}.
\label{LGU}
\end{equation}
This system has fewer equations ($n_l$) than unknowns ($n_v-n_c$). It therefore has an infinite number of possible solutions that constitute a space with the dimension $n_v-n_c-n_l$. Let $\mitbf{K}_{L^kG^k}$ be a $(n_v-n_c)\times(n_v-n_c-n_l)$ matrix containing the basis vectors of the kernel of $\mitbf{L}^k\,\mitbf{G}^k$. The general solution of (\ref{LGU}) is therefore
\begin{equation}
\hat{\mitbf{U}}^k=\mitbf{K}_{L^kG^k} \overline{\mitbf{U}}^k,
\label{Kernel}
\end{equation}
where the $n_v-n_c-n_l$ components of $\overline{\mitbf{U}}^k$ are real numbers. Injecting (\ref{Kernel}) into (\ref{MatG}) gives
\begin{equation}
\mitbf{U}^k=\mitbf{G}^k\mitbf{K}_{L^kG^k}\overline{\mitbf{U}}^k.
\label{SVD}
\end{equation}
The computation of $\mitbf{K}_{L^kG^k}$ is a key point. For this purpose, we use a classical linear algebra tool: singular value decomposition of $\mitbf{L}^k\,\mitbf{G}^k$. Technical details can be found in the Appendix A of \citet{LOMBARD04}.

\subsection{Computation of fictitious values}\label{SubSecExtra2}

Let us now consider the $n_p$ grid points of $\Omega$ in the circle ${\cal C}$ centred at $P$ with a radius $d$; for instance, $n_p=8$ in Figure \ref{FigSolMod}. At these points, we write the $k$-th order Taylor expansion in space of the solution at $P$, and then we use the expression (\ref{SVD}). This gives
\begin{equation}
\begin{array}{lll}
\mitbf{U}(x_i,\,z_j,\,t_n)&=& \displaystyle \mitbf{\Pi}_{i,j}^k\mitbf{U}^k+\mitbf{O}(h^{k+1}),\\
&&\\
&=&\displaystyle \mitbf{\Pi}_{i,j}^k\mitbf{G}^k\mitbf{K}_{L^kG^k}\overline{\mitbf{U}}^k+\mitbf{O}(h^{k+1}).
\end{array}
\label{Taylor}
\end{equation}
The set of $n_p$ equations (\ref{Taylor}) is written compactly via a $5\,n_p\times (n_v-n_c-n_l)$ matrix $\mitbf{M}$
\begin{equation}
\left(\mitbf{U}(.,\,t_n)\right)_{\cal C}=\mitbf{M}\overline{\mitbf{U}}^k+\mitbf{O}(h^{k+1}),
\label{MatM}
\end{equation}
where $\left(\mitbf{U}(.,\,t_n)\right)_{\cal C}$ is the vector containing the exact values of the solution at the grid nodes of $\Omega$ inside ${\cal C}$. These exact values are replaced by the known numerical values $\left(\mitbf{U}^n\right)_{\cal C}$, and Taylor rests are removed. From now on, numerical values and exact values of the fields are used indiscriminately. The discrete system thus obtained is overdetermined (see the remark (i) about $d$ and typical values of $n_p$ in subsection \ref{SubSecOverview}). We now compute its least-squares solution
\begin{equation}
\overline{\mitbf{U}}^k=\mitbf{M}^{-1}\left(\mitbf{U}^n\right)_{\cal C},
\label{Trace}
\end{equation}
where the $(n_v-n_c-n_l)\times 5\,n_p$ matrix $\mitbf{M}^{-1}$ denotes the pseudo-inverse of $\mitbf{M}$. From (\ref{SolMod0}), (\ref{SVD}) and (\ref{Trace}), the fictitious value in the vacuum at $(x_I,\,z_J)$ is
\begin{equation}
\begin{array}{lll}
\mitbf{U}_{I,J}^*&=& \displaystyle \mitbf{\Pi}_{I,J}^k\,\mitbf{G}^k\mitbf{K}_{L^kG^k}\mitbf{M}^{-1}\left(\mitbf{U}^n\right)_{\cal C}\\
&&\\
&=& \displaystyle \mitbf{\Lambda}_{I,J}\left(\mitbf{U}^n\right)_{\cal C}.
\end{array}
\label{Extrapolator}
\end{equation}
The $5\times 5\,n_p$ matrix $\mitbf{\Lambda}_{I,J}$ is called the \textit{extrapolator} at $(x_I,\,z_J)$. The fictitious values have no clear physical meaning. They only allow, by interpolation with numerical values inside $\Omega$, to recover the high-order Dirichlet conditions (\ref{Uk}).

\subsection{Comments and practical details}\label{SubSecOverview}

The extrapolation method described in section \ref{SubSecExtra2} has to be applied at each grid point $(I,\,J)$ in the vacuum where a fictitious value is required for the time-marching procedure. Useful comments are proposed about this method:

\begin{enumerate}
\item The radius $d$ of ${\cal C}$ must ensure that the number of equations in (\ref{MatM}) is greater than the number of unknowns: 
\begin{equation}
\varepsilon(k,\,d)=\frac{\textstyle 5\,n_p}{\textstyle n_v-n_c-n_l}\geq 1.
\label{Theta}
\end{equation}
No theoretical results are available about the optimal value of $\varepsilon$. However, numerical studies have shown that a definite overestimation ensures long-term stability: typically, $\varepsilon \approx 4$. Various strategies can be used to ensure (\ref{Theta}), such as an adaptative choice of $d$ depending on the local geometry of $\Gamma$ at $P$. Here we adopt a simpler strategy consisting in using a constant radius $d$. With $k=3$, numerical experiments have shown that $d=3.2\,h$ is a good candidate for this purpose. In this case, one typically obtains  $n_p \approx 15$.

\item  Since the boundary conditions do not vary with time, the extrapolators $\mitbf{\Lambda}_{I,J}$ in (\ref{Extrapolator}) can be computed and stored during a pre-processing step. At each time step, only small matrix-vector products are required. 

\item The extrapolators $\mitbf{\Lambda}_{I,J}$ account for the local geometry of $\Gamma$ at the projection points $P$ on $\Gamma$ via $\mitbf{L}^k$ (section \ref{SubSecBC}). Moreover, they incorporate the position of $P$ relative to the Cartesian meshing, via $\mitbf{\Pi}_{i,j}$ (\ref{Taylor}) and $\mitbf{\Pi}_{I,J}$ (\ref{Extrapolator}). The set of extrapolators therefore provides a subcell resolution of $\Gamma$ in the meshing, avoiding the spurious diffractions induced by a naive description of the boundaries.

\item The stability of the method has not been proved. However, numerical experiments clearly indicate that the CFL condition of stability is not modified compared with the case of a homogeneous medium. The solution does not grow with time, even in the case of long-time simulations (see section \ref{SubSecTest4}).

\item In a previous one-dimensional study \citep{PIRAUX01}, the local truncation error of the method has been rigorously analysed, leading to the following result: using the fictitious values (\ref{Extrapolator}) ensures a local $r$-th order spatial accuracy if $k\geq r$, where $r$ is the order of spatial accuracy of the scheme. In 2D configrations with material interfaces \citep{LOMBARD04,LOMBARD06}, no proof has been conducted, but numerical experiments have shown that the $r$-th order overall accuracy is also maintained by taking $k=r$. Note that a slightly smaller order of extrapolation can be used: $k=r-1$ suffices to provide $r$-th order overall accuracy \citep{GUSTAFSSON75}. The value $k=3$ is therefore used for the fourth-order ADER scheme.

\item The extrapolators do not depend on the numerical scheme adopted. They depend only on $k$ and on physical and geometrical features. Standard subroutines for computing the extrapolators $\mitbf{\Lambda}_{I,J}$ can therefore be developed and adapted to a wide range of schemes. Subroutines of this kind are freely available in FORTRAN at the web page {\tt http://w3lma.cnrs-mrs.fr/\textasciitilde MI/Software/}. 
\end{enumerate}

\subsection{Case of staggered-grid schemes}\label{SubSecStag}

\begin{figure}
\begin{center}
\begin{tabular}{cc}
(a) & (b)\\
\includegraphics[scale=0.56]{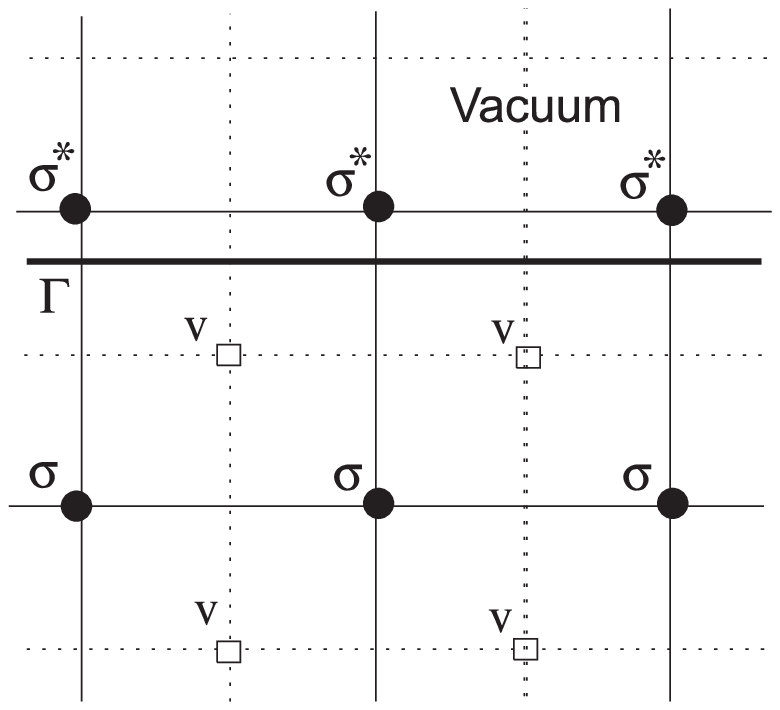}&
\includegraphics[scale=0.56]{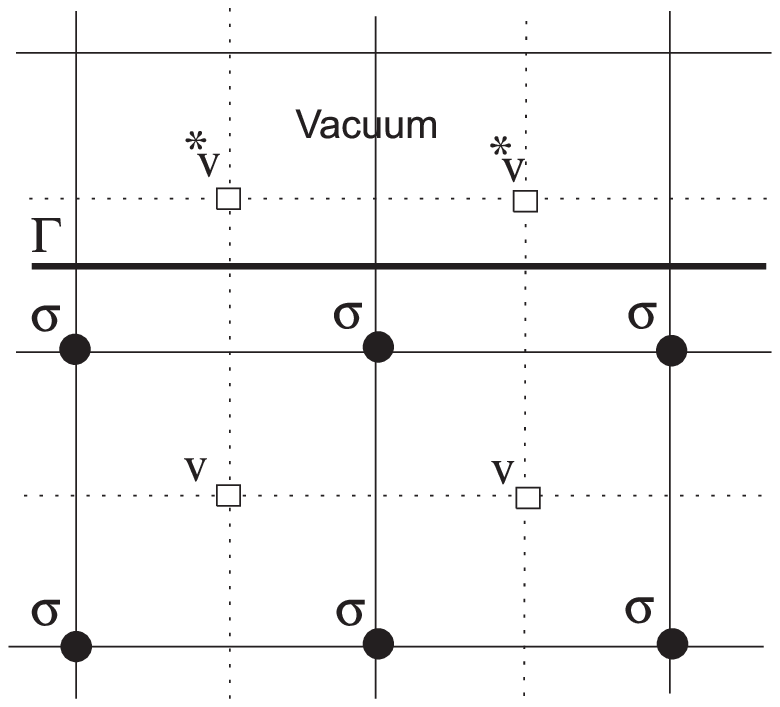}
\end{tabular}
\end{center}
\caption{\textit{Staggered-grid schemes with a plane boundary $\Gamma$ parallel to the meshing: two cases can be distinguished, depending on the position of $\Gamma$ relative to the meshing. Case (a), where the fictitious stress is estimated, works well, while case (b), where the fictitious velocity is estimated, leads to long-term instabilities.}}
\label{FigStag}
\end{figure} 

Instead of using a single-grid scheme as proposed in section \ref{SubSecScheme}, readers may be interested in adapting our approach to staggered-grid schemes such as CSS or PSS (see section \ref{SecIntro} for the definition of these terms). However, in the case of some of the boundary positions relative to the meshing, computational instabilities occur, especially when long-time integration is considered.

To understand why this is so, let us consider PSS-2. Taking a simple flat boundary to exist between the medium and the vacuum leads to two typical geometrical configurations. At one position of the free surface, the boundary discretization will require only the stress field to be extrapolated (Figure \ref{FigStag}-(a)). Our procedure works satisfactorily with this type of discretization at any order $k$. It also yields stable and accurate solutions when dealing with PSS-4, contrary to the vacuum method. Using 10 grid nodes per minimal S-wavelength gives similar performance in this case to those of our numerical experiments based on the ADER scheme, which are shown in section \ref{SecNum}.

At another position of the free surface where only extrapolated velocities are required within a wide zone (Figure \ref{FigStag}-(b)), our procedure results in instabilities. The reason for this problem is as follows: fictitious velocities involve first-order boundary conditions (\ref{L0t}) and higher-order conditions (see section \ref{SubSecBC}), but they do not involve the fundamental zeroth-order Dirichlet conditions (\ref{L0U0}). Since the latter conditions are never enforced, an increasing oscillating drift occurs near the boundary, which invalidates the computations. Similar behavior is observed with PSS-4, but after a longer time: the numerical solution generally works  well during a few thousand time steps, before growing in a unstable manner. 

The extrapolation method presented here is therefore not recommended for use with staggered-grid schemes, especially PSS-2, except in the trivial case sketched in Figure \ref{FigStag}-(a).

\begin{figure}
\begin{center}
\includegraphics[scale=0.7]{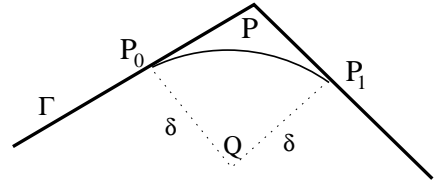}
\caption{{\it Boundary $\Gamma$ with a corner at $P$, replaced locally by an arc of circle with a radius $\delta$ between $P_0$ and $P_1$}.}
\label{FigCorner}
\end{center}
\end{figure} 

\begin{figure}
\begin{center}
\begin{tabular}{c}
(a)\\
\includegraphics[scale=0.36]{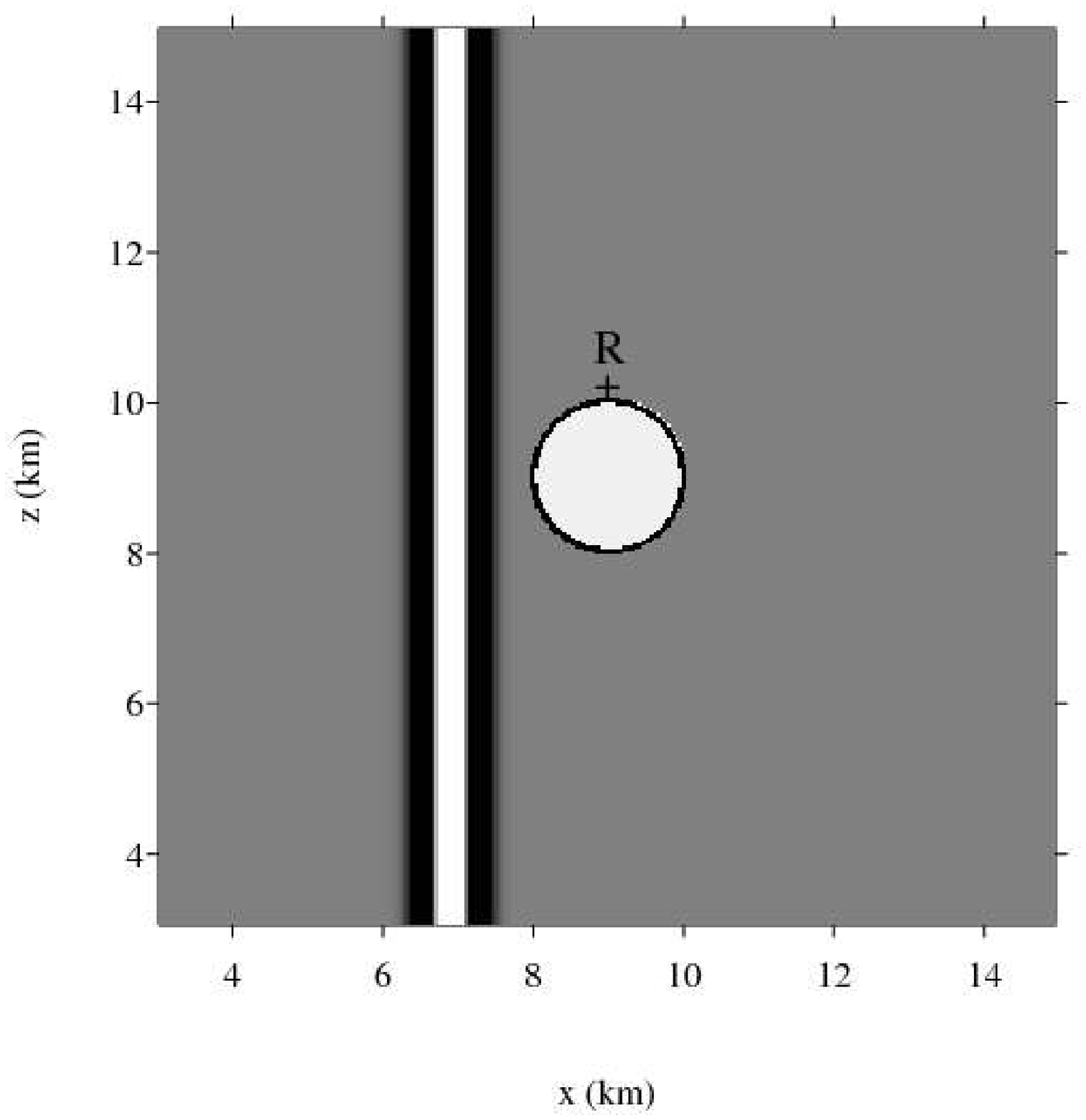}\\
(b)\\
\includegraphics[scale=0.36]{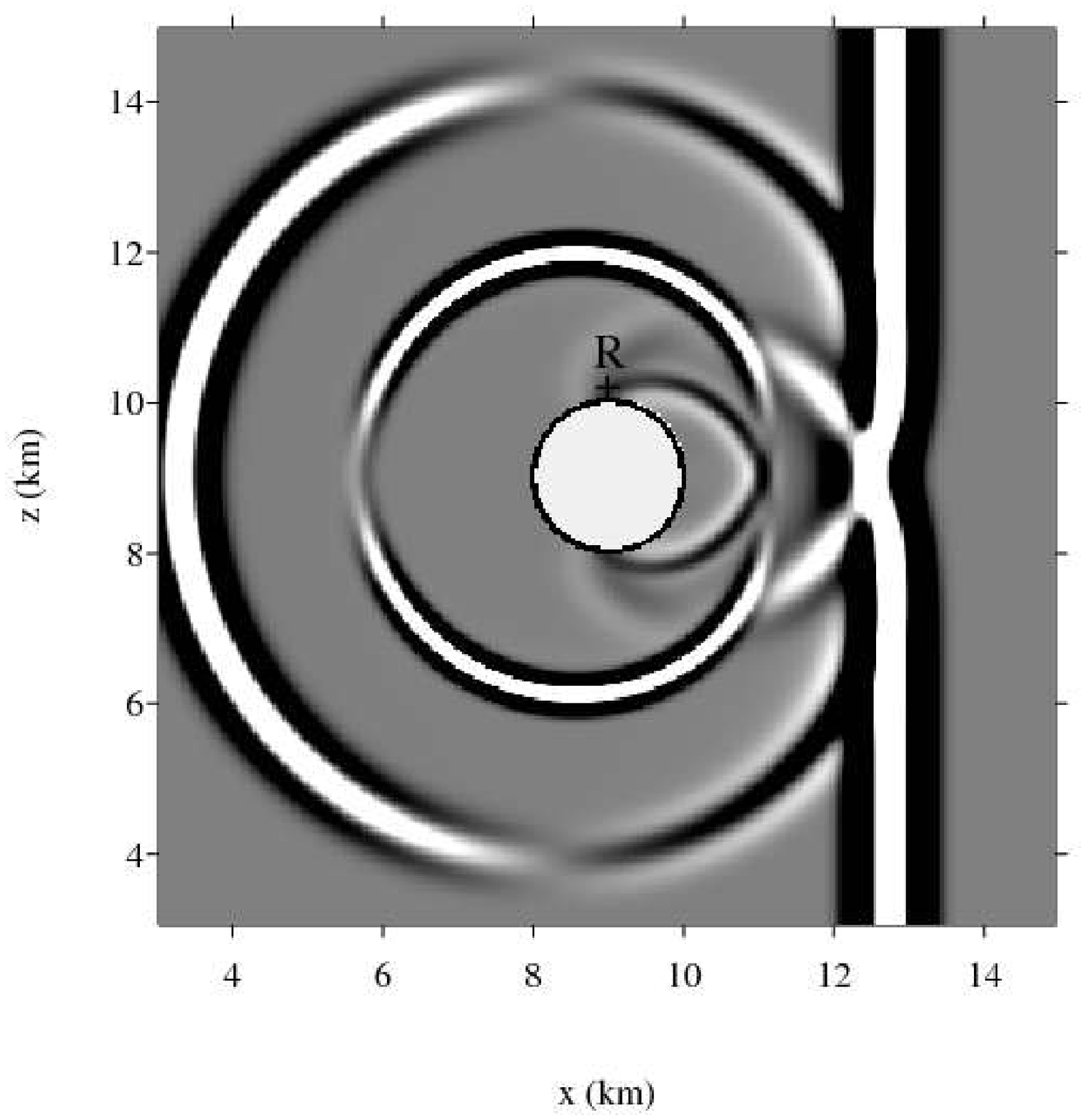}\\
(c)\\
\includegraphics[scale=0.36]{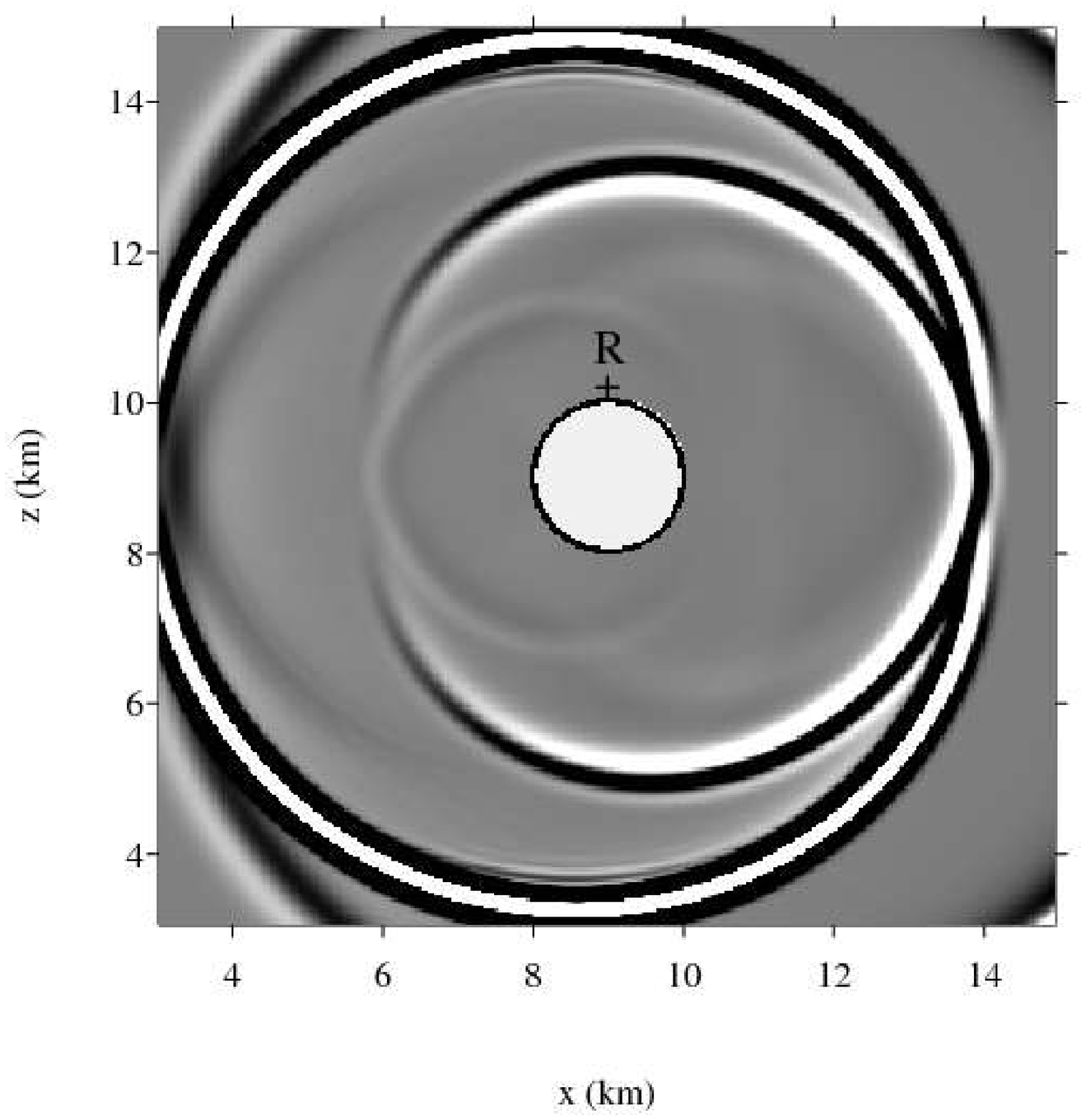}
\end{tabular}
\end{center}
\caption{\textit{Test 1: snapshots of $v_x$ at the initial instant (a), at mid-term (b) and at the final instant (c).}}
\label{FigTest1A}
\end{figure} 

\begin{figure}
\begin{center}
\begin{tabular}{c}
(a)\\
\includegraphics[scale=0.325]{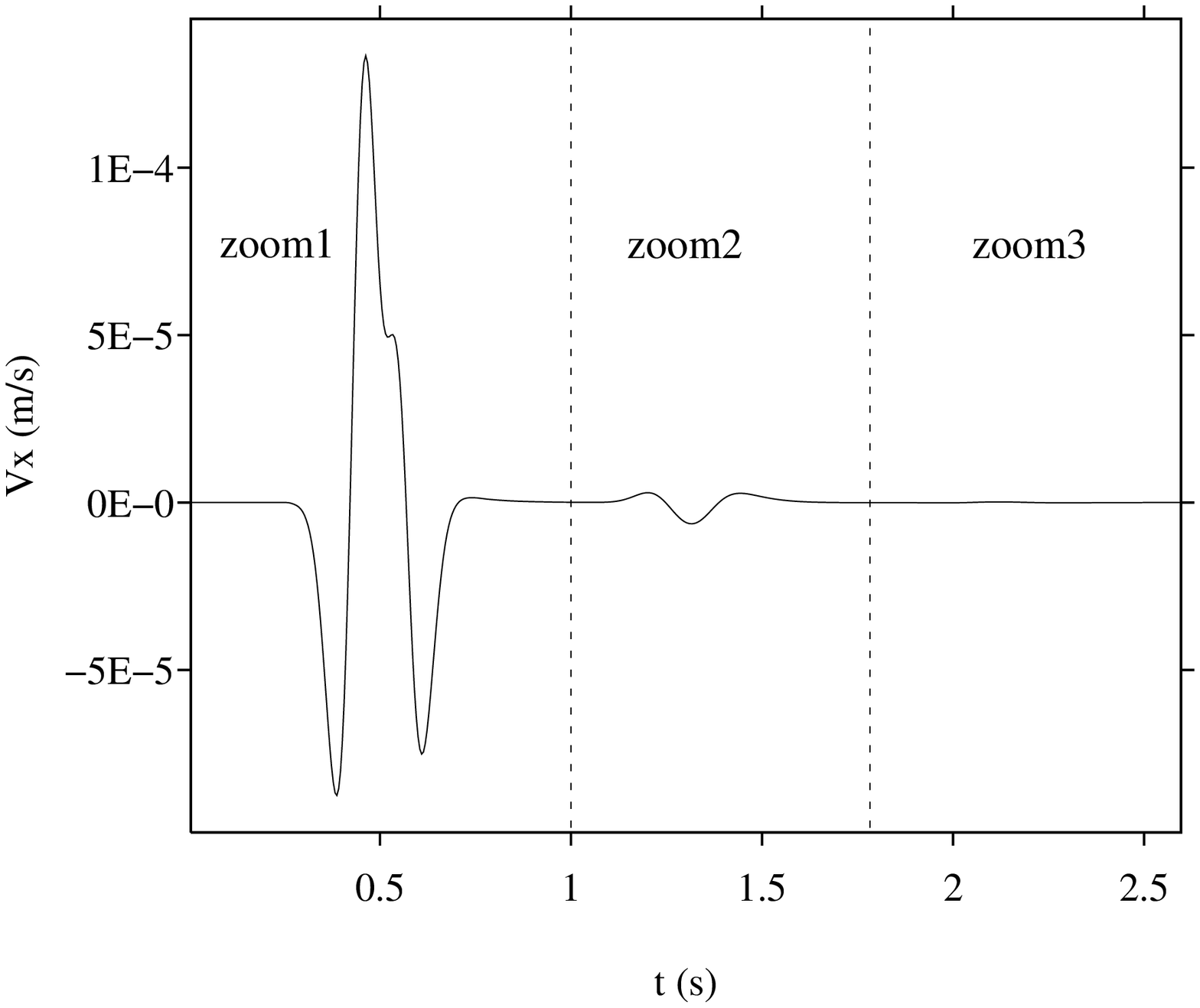}\\
(b)\\
\includegraphics[scale=0.325]{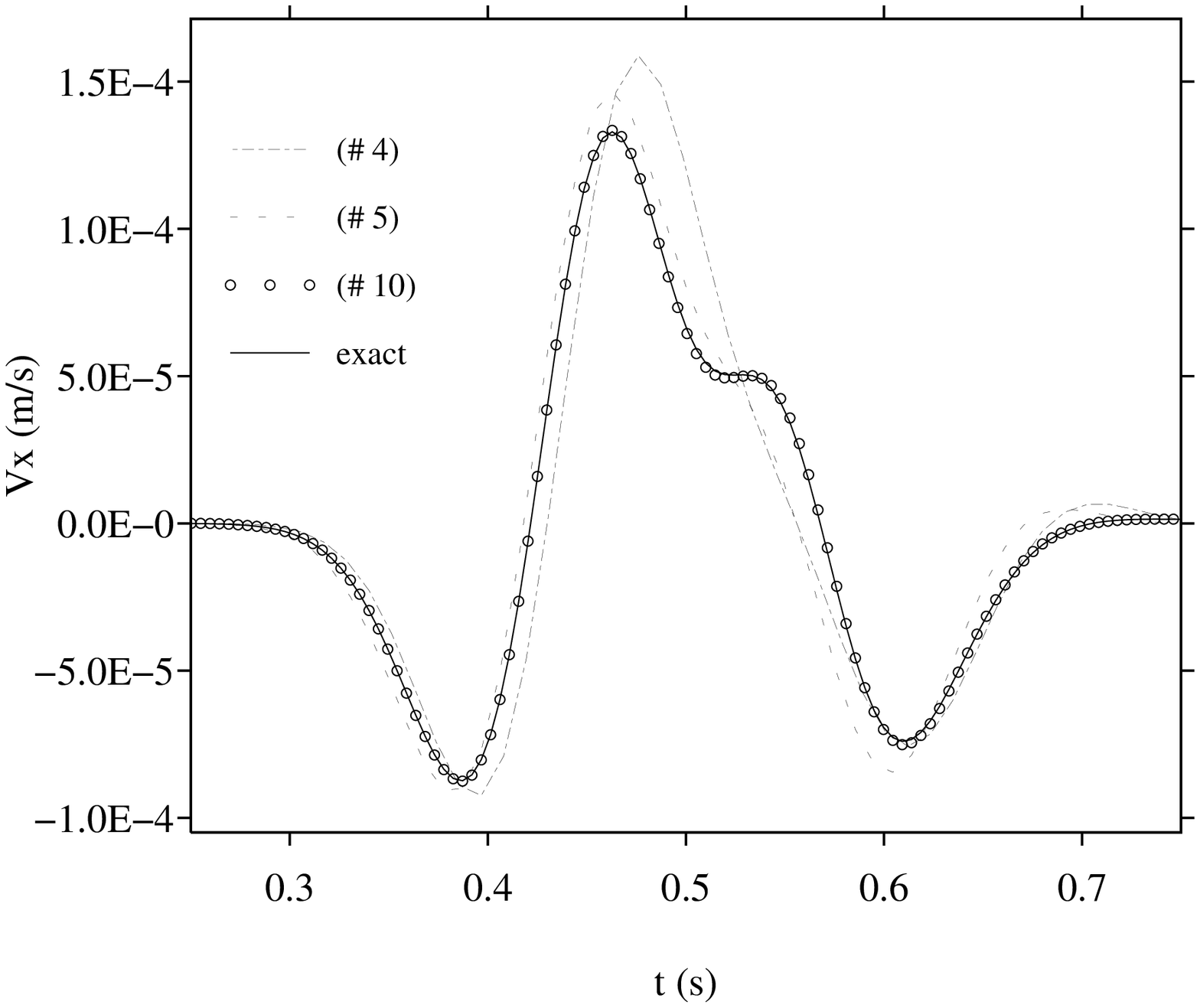}\\
(c)\\
\includegraphics[scale=0.325]{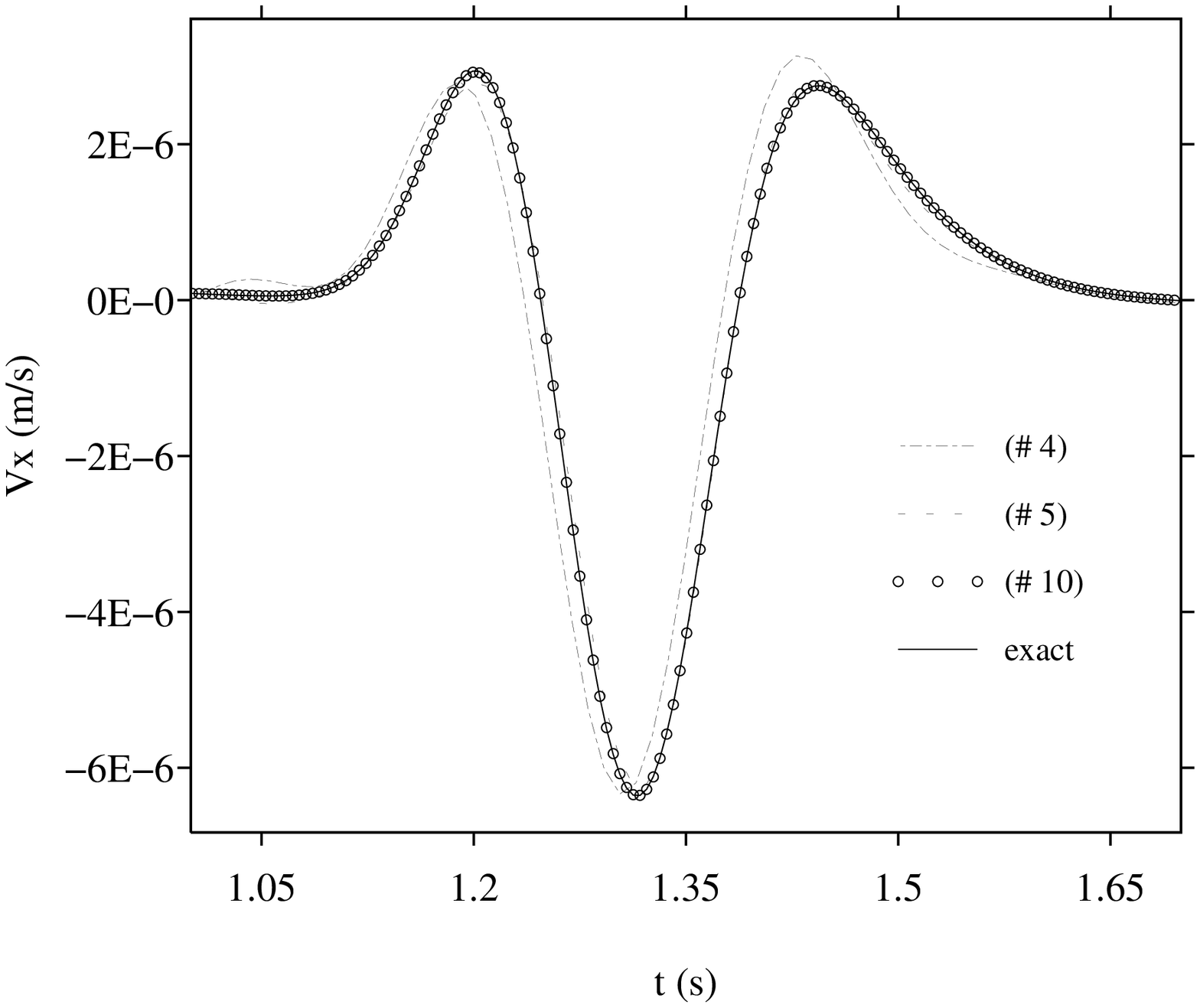}\\
(d)\\
\includegraphics[scale=0.325]{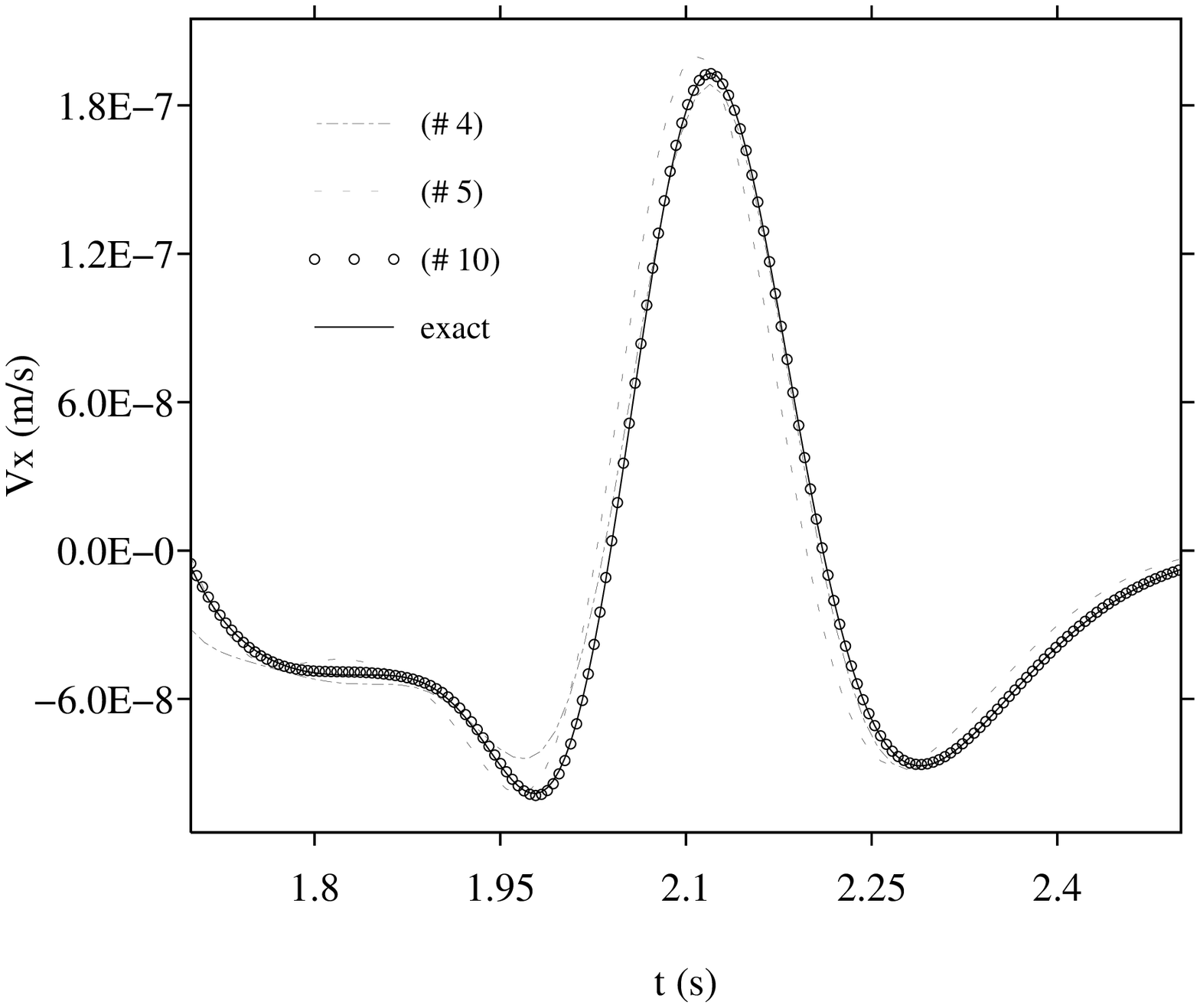}
\end{tabular}
\end{center}
\caption{\textit{Test 1: time history of $v_x$ (a). Zooms on successive time windows, with various discretizations (b,c,d): the number after \# denotes the number of grid nodes per minimal S-wavelength.}}
\label{FigTest1B}
\end{figure} 

\subsection{Case of non-smooth geometries}

Up to know, we have assumed that the boundary $\Gamma$ was sufficiently smooth at the projection points, being at least $C^{k+1}$ at each $P$, where $k\geq 0$ is the order of differentiation defined in section \ref{SubSecOverview}. Let us assume now that $\Gamma$ is only $C^K$ at a point $P$, with $K<k+1$. Then, the components of $\mitbf{L}^k$ in (\ref{LkUk}) involving the derivatives $\frac{d^\alpha\,x}{d\,x^\alpha}(\tau)$ and $\frac{d^\alpha\,z}{d\,x^\alpha}(\tau)$ ($\alpha=K+1,...,k+1$) of the parametric representation are discontinuous, invalidating locally the method proposed. In our software, we have implemented the following rough treatment:
\begin{enumerate}
\item If $K=0$, the boundary owns a corner and the solution has an integrable singularity. The corner is replaced by an arc of circle centred at $Q$ with radius $\delta$ (figure \ref{FigCorner}), leading to a $C^1$ curve.
\item If $0<K<k+1$, as in the previous case at $P_0$ and $P_1$ with $K=1$, the values of $\frac{d^\alpha\,x}{d\,x^\alpha}(\tau)$ and $\frac{d^\alpha\,z}{d\,x^\alpha}(\tau)$ ($\alpha=K+1,...,k+1$) are taken indiscriminately on one side or the other of the point considered. 
\end{enumerate}
No numerical instabilities were observed if $\delta$ (in case (i)) or the radius of curvature (in case (ii)) is much greater than $h$. It is agreed that the accuracy of computations is no more controlled in the cases (i) and (ii), especially the convergence towards the exact solution. 

More sophisticated treatments of geometrical singularities, such as space-time mesh refinement \citep{BERGER98}, require further investigation, which is out of the scope of the present paper. New studies are also needed in the case of merging boundaries, occuring for instance when an internal material interface reaches the free surface \citep{MOCZO04}.

\section{Numerical experiments}\label{SecNum}

\subsection{Configurations}

The time evolution of the source is a Ricker wavelet
\begin{equation}
g(t)=\left\{2\left(\pi\,f_c\left(t-t_c\right)\right)^2-1\right\}e^{\textstyle -\left(\pi\,f_c\left(t-t_c\right)\right)^2},
\label{RICKER}
\end{equation}
where $f_c$ is the central frequency, and $t_c=1/f_c$. The maximal frequency $f_{\max}$ defined by $|\tilde{g}(f_{\max})/\tilde{g}(f_c)|=0.5$ (the tilde designates the Fourier transform) is $f_{\max}\approx 1.6\,f_c$. We will adopt this frequency $f_{\max}$ for our rule of thumb about
the number of grid nodes per S-wavelength. The following values of the physical parameters will be used in all the following tests: $\rho=2400\,\mbox{kg/m}^3$, $c_p=4500$ m/s, and $c_s=2200$ m/s. Lastly, the mesh size and the time step satisfy $c_p\,\Delta\,t/h= 0.85$. 

The simulations are performed on a PC Pentium at 3 GHz with 2 GB of RAM. The results of tests 1 and 2, with constant and null curvature of $\Gamma$, compare quantitatively with analytical solutions denoted by a solid line. Test 3, with a variable curvature, is purely qualitative. Test 4 shows the slow decrease in the mechanical energy which occurs during very long integration times, which confirms the stability of the method. 

\subsection{Test 1: circular boundary}\label{SubSecTest1}

\begin{figure}
\begin{center}
\begin{tabular}{c}
(a)\\
\includegraphics[scale=0.42]{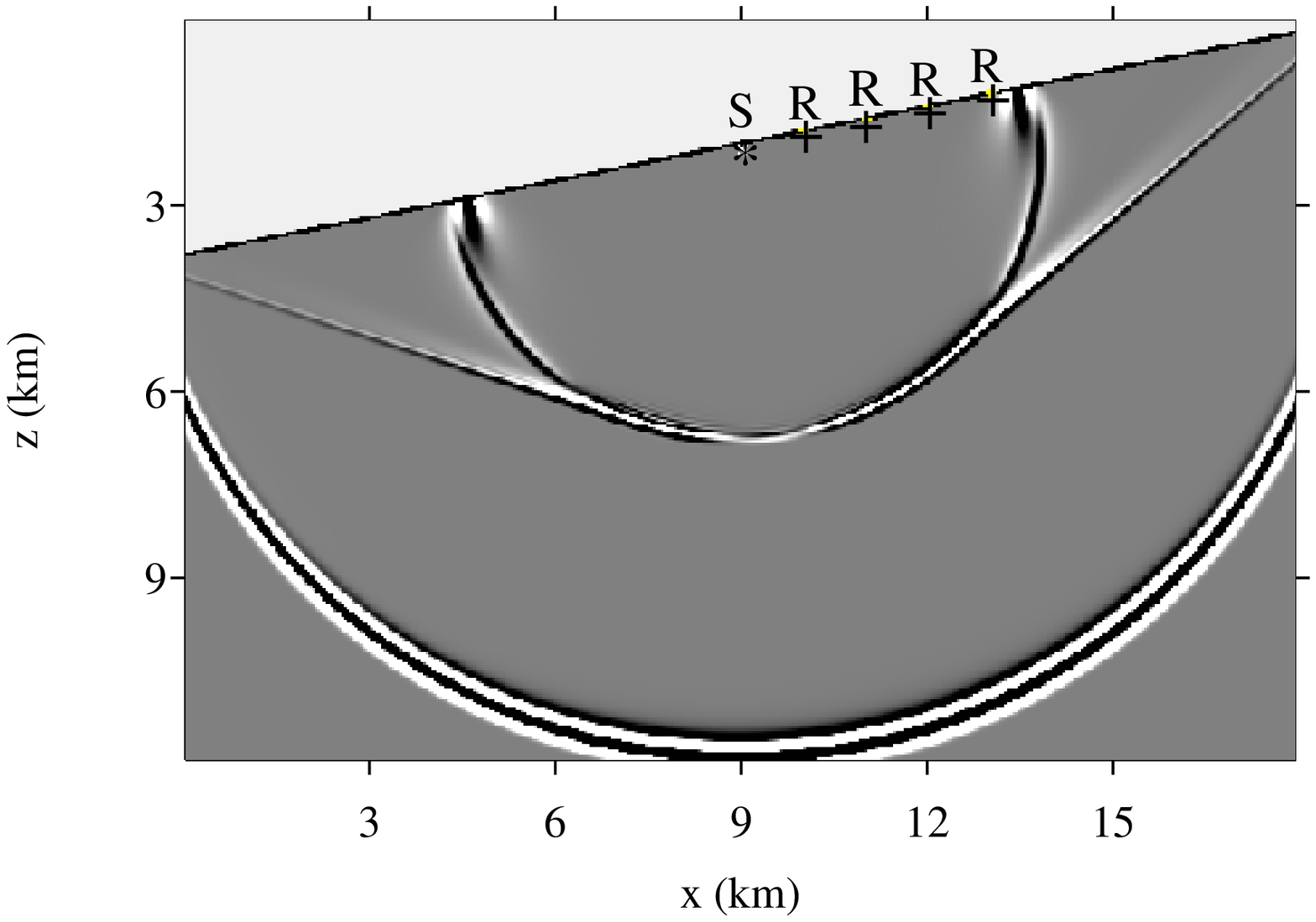}\\
(b)\\
\includegraphics[scale=0.42]{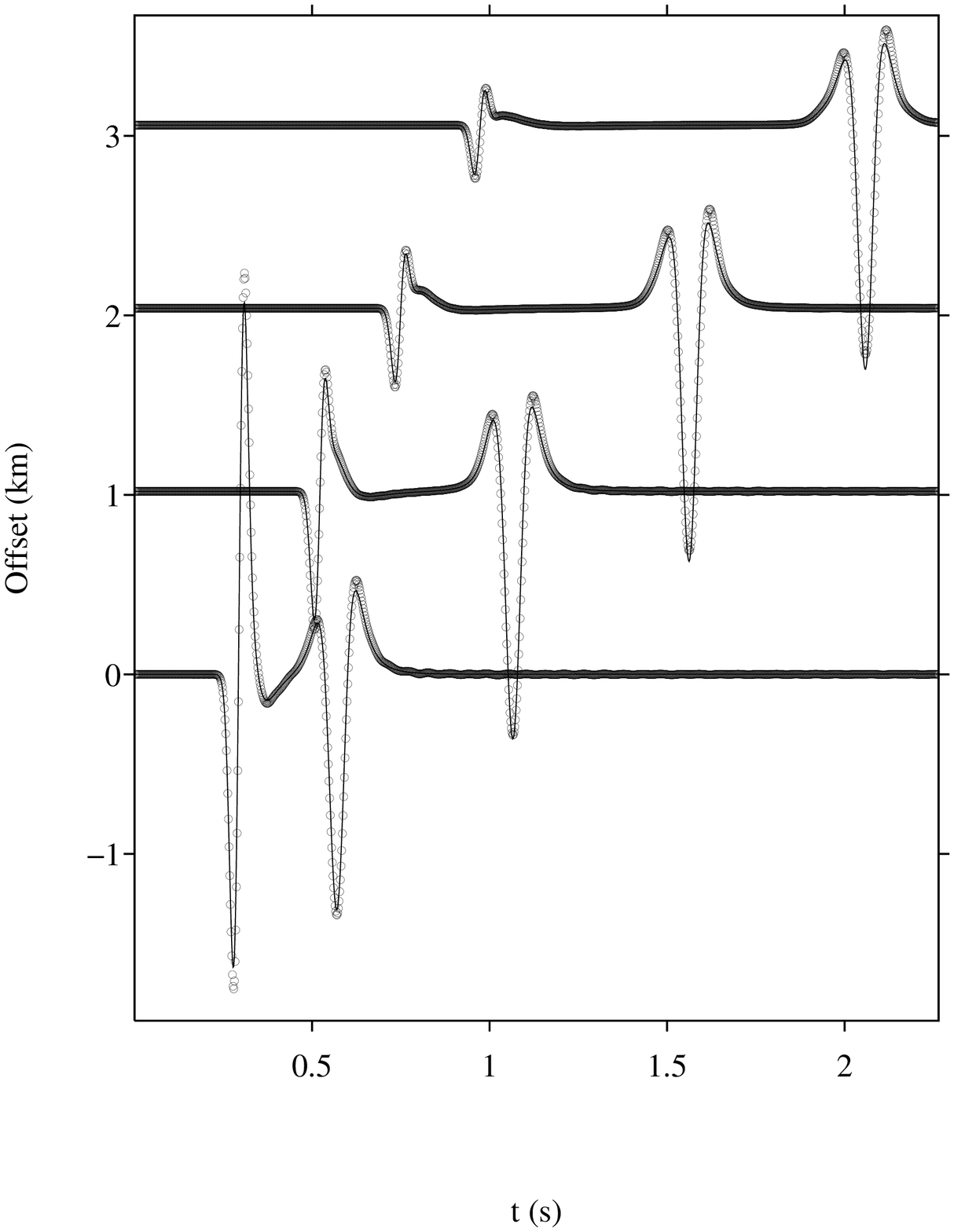}
\end{tabular}
\end{center}
\caption{\textit{Test 2: snapshot of $v_z$ at the final instant (a). Numerical and exact time histories of $v_z$ (b)}.}
\label{FigTest2}
\end{figure} 

\begin{figure}
\begin{center}
\begin{tabular}{c}
\includegraphics[scale=0.4]{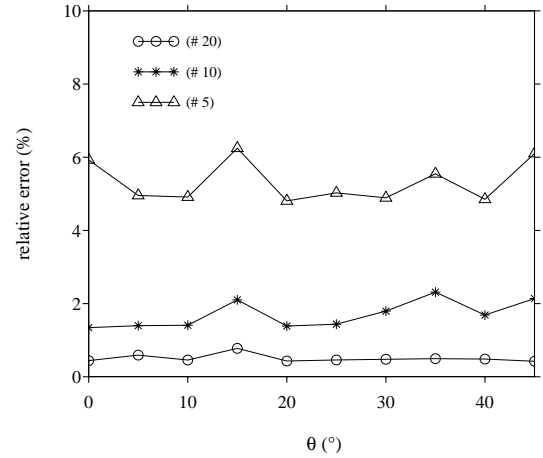}
\end{tabular}
\end{center}
\caption{\textit{Test 2: parametric study of the relative error in terms of the boundary's angle $\theta$, with various discretizations. The number after \# denotes the number of grid nodes per minimal S-wavelength.}}
\label{Test2Param}
\end{figure} 

\textit{Computational efficiency}. Let us consider a circular cavity containing vacuum, with radius 1 km, at the center of a 18 km $\times$ 18 km domain. In a first part, the mesh spacing is $h=25$ m. The source is a rightward-moving plane wave, with $f_{\max}=8$ Hz, ensuring 22 grid nodes per minimal P-wavelength and 10 grid nodes per minimal S-wavelength at that frequency. 

During the pre-processing step, the program finds the 616 grid nodes where fictitious values are required; it also computes and stores the 616 extrapolators defined by the expression (\ref{Extrapolator}). Time integration is performed in 550 time steps, which corresponds to a propagation time of 2.75 s and a propagation distance of 22 minimal wavelengths. The preprocessing step takes 21 s of CPU time. The time integration takes 1100 s of CPU time, including 28.6 s induced by the computation and by the use of fictitious values, which amounts to an extra time cost of only 2.6 \%. Figure \ref{FigTest1A} shows snapshots of $v_x$ at the initial instant (a), after 275 time steps (b) and after 550 time steps (c). 

\textit{Quantitative study}. In a second part, three discretizations are considered: $h=25$ m, 50 m, 60 m, corresponding respectively to 10, 5 and 4 grid nodes per minimal S-wavelength. A receiver is set just above the cavity, at the position (9 km, 10.2 km), that can be seen on Figure \ref{FigTest1A}; it mainly records the waves propagating along the boundary. Numerically, these waves are highly sensitive to the quality of the fictitious values defined in section \ref{SecDiscrete}. 

Figure \ref{FigTest1B}-(a) shows the time history of $v_x$ at the receiver. In this time window, three main wave packets are generated; with the scale of Figure \ref{FigTest1B}-(a), the third packet cannot be seen. The amplitude is divided by a factor of approximately 30 from one packet to the following one. For the sake of clarity, zooms around each wave packet are shown in Figure \ref{FigTest1B}-(b,c,d). These solutions are compared with an exact solution computed thanks to inverse Fourier transforms on 4096 frequencies, with $1.25\,10^{-2}$ Hz as the sampling frequency; each harmonic component is expanded into 60 Bessel modes. The agreement between the numerical and the analytical values is very good when 10 grid nodes per wavelength are used, even at very small amplitudes (d). For 5 and 4 grid nodes per wavelength, the solution is slightly less accurate, but it is still acceptable. 

\subsection{Test 2: inclined straight boundary}\label{SubSecTest2}

\begin{figure}
\begin{center}
\begin{tabular}{c}
(a)\\
\includegraphics[scale=0.4]{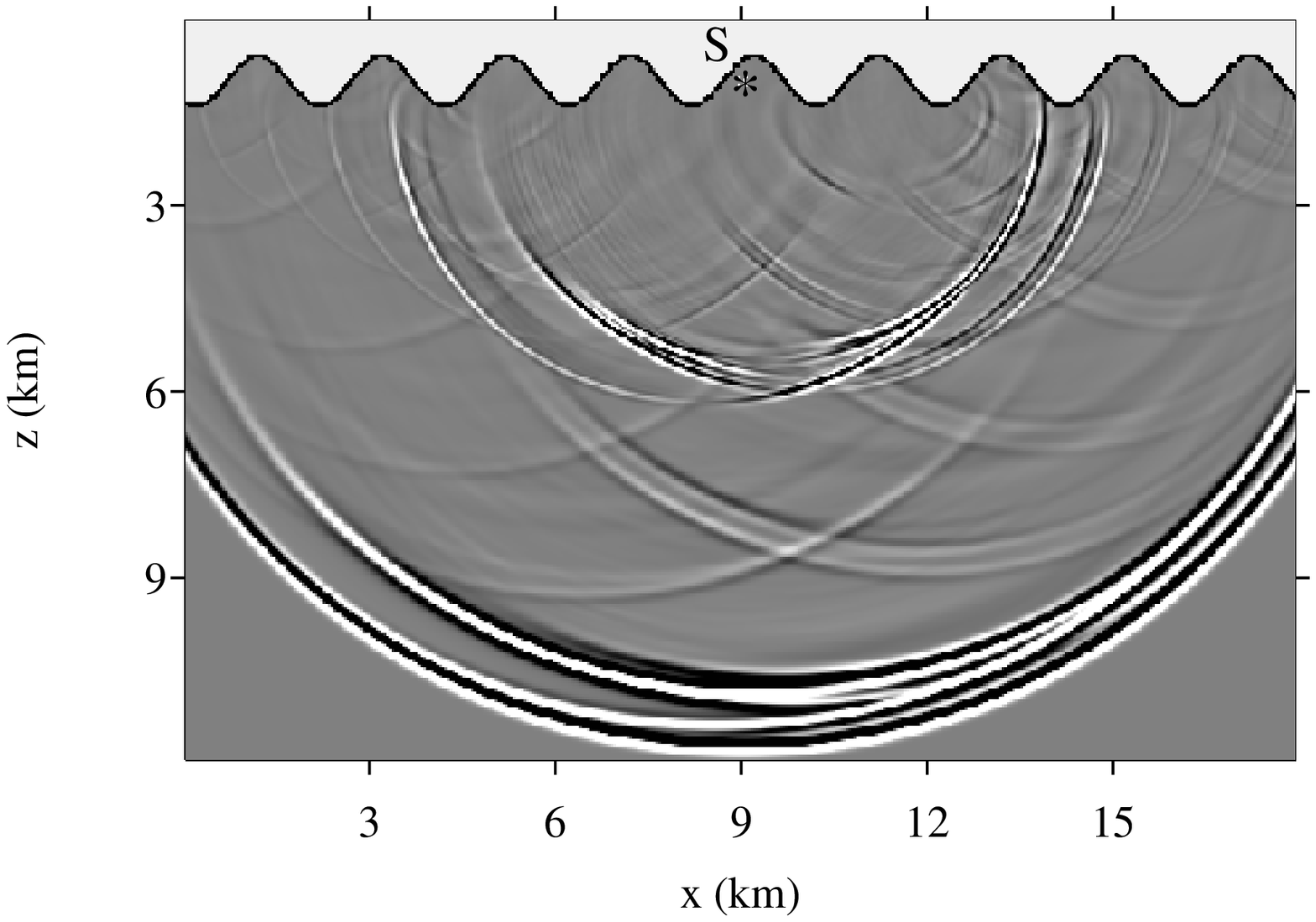}\\
(b)\\
\includegraphics[scale=0.4]{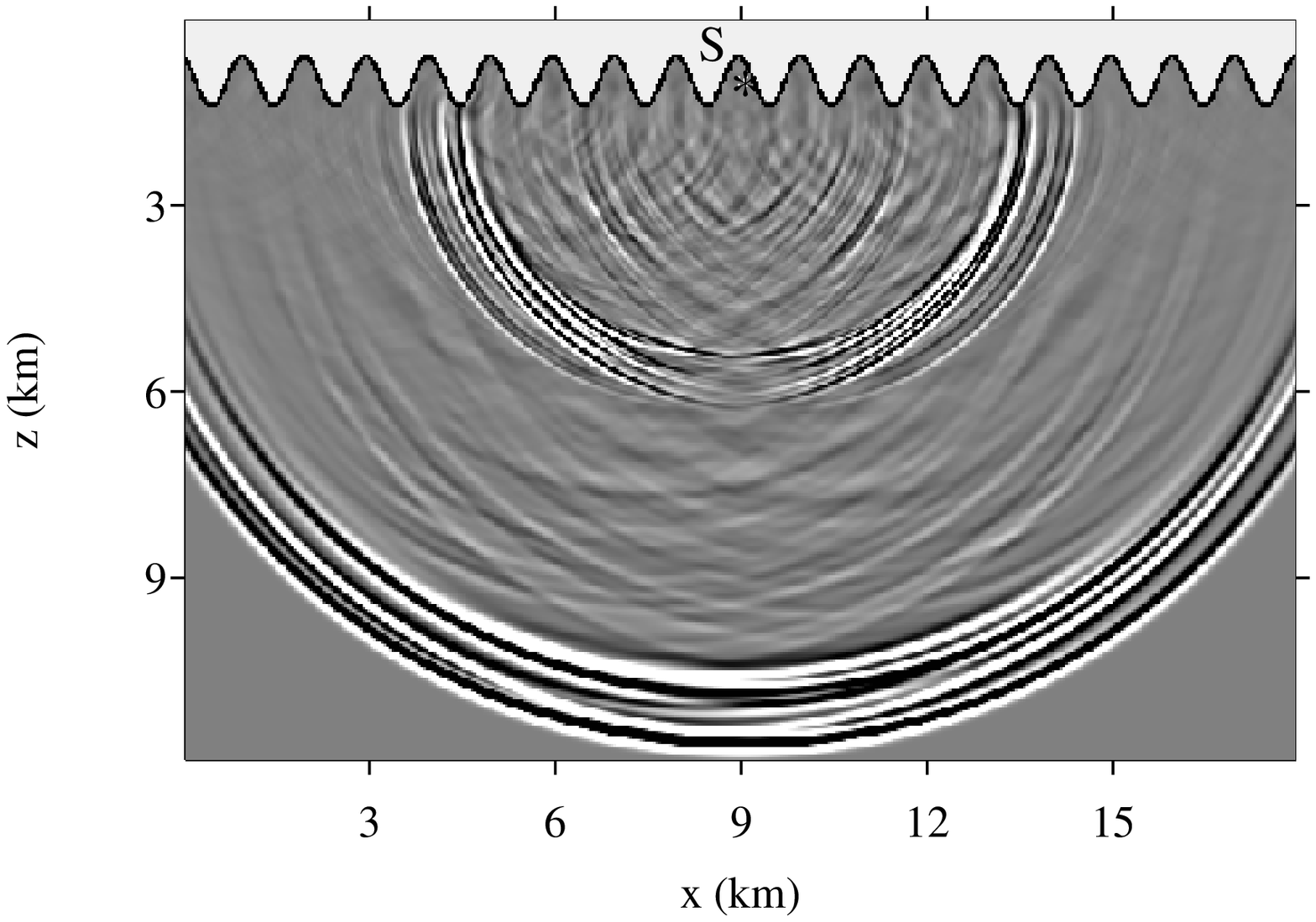}\\
(c)\\
\includegraphics[scale=0.4]{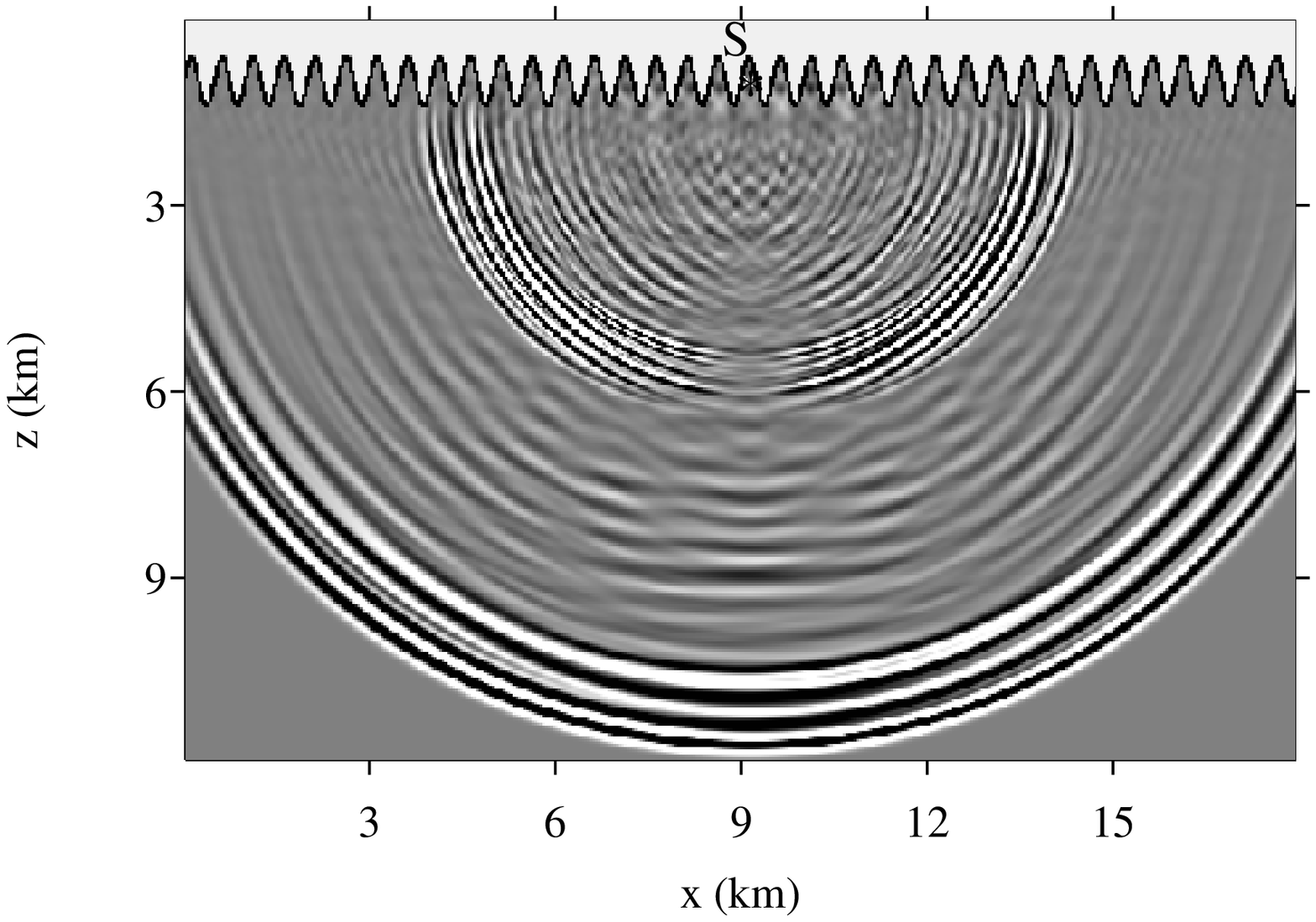}
\end{tabular}
\end{center}
\caption{\textit{Test 3: snapshots of $v_z$, for various sinusoidal topographies}.}
\label{FigTest3}
\end{figure} 

\begin{figure}
\begin{center}
\begin{tabular}{c}
(a)\\
\includegraphics[scale=0.43]{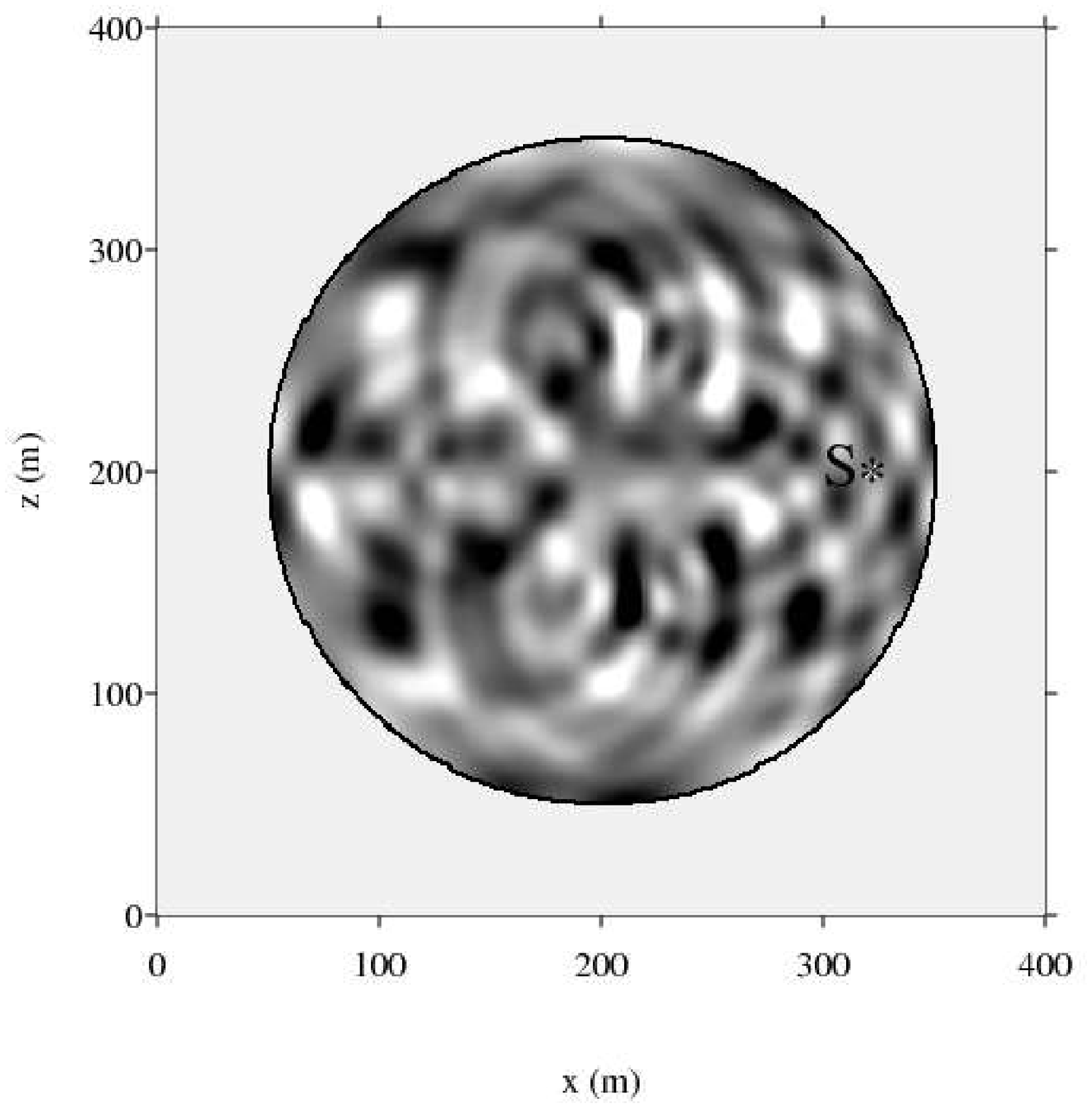}\\
(b)\\
\includegraphics[scale=0.43]{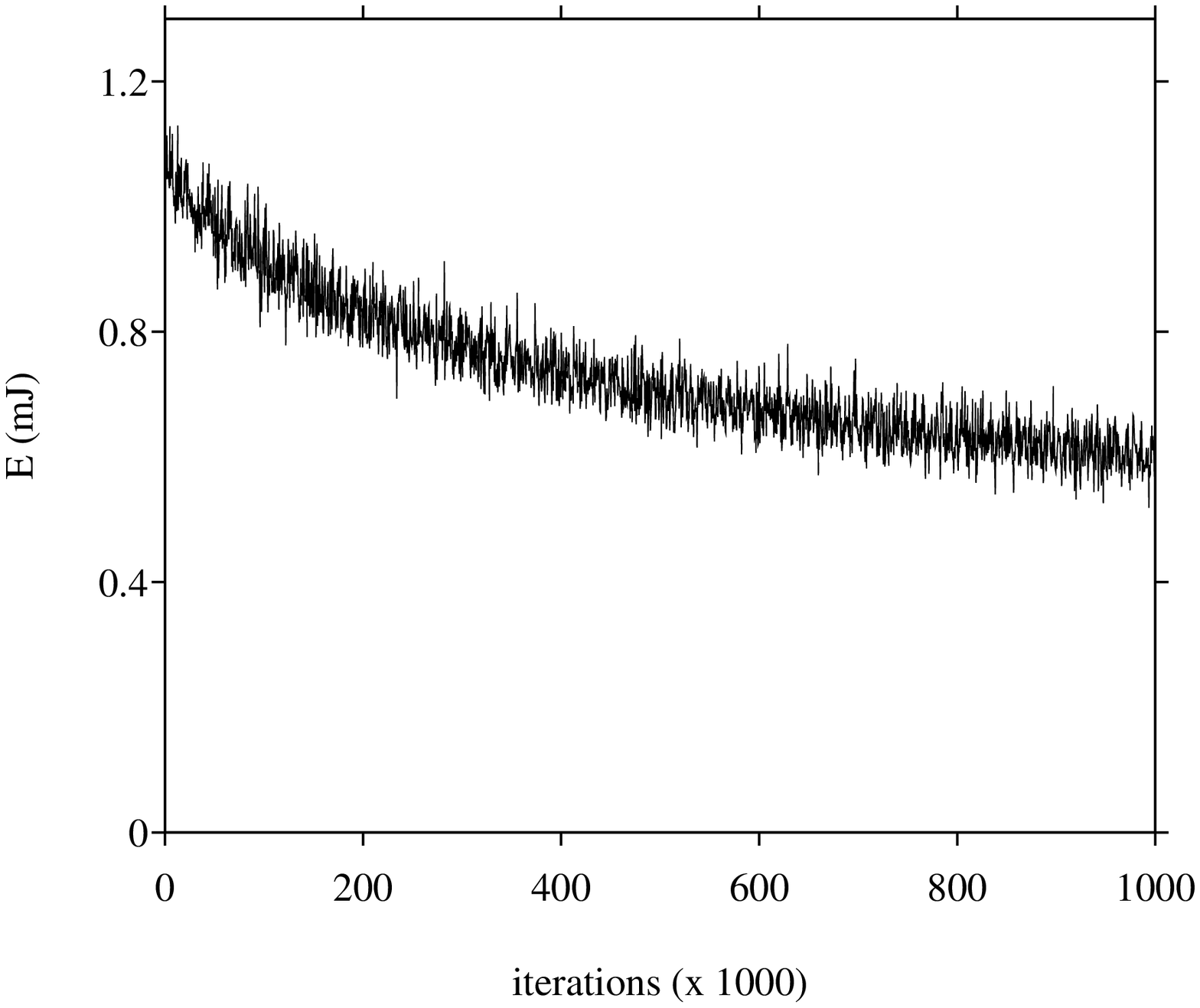}
\end{tabular}
\end{center}
\caption{\textit{Test 4: snapshot of $v_z$ (a) and time history of the mechanical energy (b)}.}
\label{FigTest4}
\end{figure} 

\textit{The Garvin's problem}. As a second test, we take a plane boundary inclined against the Cartesian mesh. The domain under investigation is 18 km wide and 12 km high, with the origin of the coordinates on the top and left. The mesh spacing is $h=10$ m. Four receivers at (10 km, 1.8 km), (11 km, 1.6 km), (12 km, 1.4 km) and (13 km, 1.2 km) belong to the free boundary which is inclined at an angle of $\theta=11.3^{^\circ}$ relative to $Ox$. 

An explosive source S is buried at (9 km, 2.1 km), with $f_{\max}=24$ Hz. The distance between the source and the free surface is roughly 100 m $<\lambda_p/3$, where $\lambda_p$ is the wavelength of the compressional waves at frequency $f_c$, and hence large Rayleigh waves are generated, with a velocity $c_r=2054$ m/s. To prevent the occurence of spurious oscillations, the source is spread out numerically over a radius of $R=\lambda_p/7.5=40$ m. The source is weighted by a gaussian law with a standard deviation $R/2=20$ m. The spatial discretization ensures the sampling of roughly 18 grid nodes per minimal P-wavelength, and 9 grid nodes per minimal S-wavelength at the frequency $f_{\max}$.

Figure \ref{FigTest2}-(a) shows a snapshot of $v_z$ after 1200 time steps, corresponding to a propagation time of 2.25 s and a propagation distance of 55 minimal wavelengths. Direct cylindrical waves are observed, together with converted PP waves, converted PS waves (with an almost linear wavefront), and Rayleigh waves. In Figure \ref{FigTest2}-(b), the time history of $v_z$ recorded at the receivers can be favourably compared with an exact solution. The latter is obtained by convolving the Green's function obtained by the well-known Cagniard-de Hoop method \citep{GARVIN55,SANCHEZ06} with the source wavelet (\ref{RICKER}) and with the discrete source spreading. 

\textit{Influence of the slope}. To quantify the effects of the angle between the boundary and the Cartesian meshing on the numerical solution, we perform a parametric study of the error in terms of $\theta$. Ten angles are considered, from $\theta=0^{\circ}$ to $\theta=45^{\circ}$ in steps of $5^{\circ}$. In each configuration, the waves are measured at the free boundary after propagating for 65 minimal wavelengths. The error of $\mitbf{v}.\mitbf{n}$ is measured in norm $L_2$, and then it is normalized by the norm $L_2$ of the exact time history of $\mitbf{v}.\mitbf{n}$.

The results of this study are shown in Figure \ref{Test2Param}, with various discretizations: 5, 10, 20 grid nodes per minimal S-wavelength. With a given $h$, the error is almost constant and independent of $\theta$. This constitutes a crucial advantage of our method over the vacuum method, where the error at $45^{\circ}$ is much greater than that at $0^{\circ}$: it means that an extremely fine discretization is required to obtain accurate results with the vacuum method when arbitrary-shaped boundaries are encountered \citep{BOHLEN06}.  

\subsection{Test 3: sinusoidal boundary}\label{SubSecTest3}

Since boundaries not related to the finite-difference grid can be included, the third test is performed on a sinusoidal free boundary, with a peak-to-peak amplitude of 800 m and various wavelengths: 0.5 km, 1 km and 2 km. The sinusoidal curve is centred around $z=1$ km.  The source S is located at (9 km, 1 km). The other parameters are the same as in test 2. Figure \ref{FigTest3} shows snapshots of $v_z$ at the final instant 2.25 s. One can clearly see how the wavelength of the sinusoidal boundary influences the diffracted fields. 

Convergence studies (not shown here) were performed in these three cases, by comparing solutions computed on finer grids. We again concluded that accurate solutions can be obtained when the simulations involve approximately 10 grid nodes per minimal S-wavelength at the frequency $f_{\max}$ of the source wavelet, even in the case of complex topographies with variable curvatures.

\subsection{Test 4: long-term stability}\label{SubSecTest4}

The fourth test focuses on long-term stability \citep{STACEY94,HESTHOLM03}. For this purpose, we consider a circular elastic domain with a radius of 150 m, surrounded by vacuum. The source S is located inside the circle, at (320 m, 200 m). This configuration is obviously not realistic, but it enlights the influence of the boundary on the numerical solution after many reflections, and especially on the possible excitation of numerical spurious modes leading to long-term instability. The mesh size is $h=1$ m. Time integration is performed during $10^6$ time steps, with $f_{\max}=160$ Hz. 

Figure \ref{FigTest4}-(a) shows a snapshot of $v_z$ at the final instant: no instability is observed, and the antisymmetry of $v_z$ is satisfied. Once the source is extincted ($t>2\,t_c$), the mechanical energy $E$ is theoretically maintained. It can be written in terms of $\mitbf{v}$ and $\mitbf{\sigma}$
\begin{equation}
\begin{array}{l}
\displaystyle
E=\frac{\textstyle 1}{\textstyle 2}\int\int_\Omega\left\{ \rho\,\mitbf{v}^2+\frac{\textstyle \lambda+2\,\mu}{\textstyle 4\,\mu\,(\lambda+\mu)}\left(\sigma_{xx}^2+\sigma_{zz}^2\right)+\frac{\textstyle 1}{\textstyle \mu}\sigma_{xz}^2\right.\\
[8pt]
\displaystyle
\hspace{2cm}\left.-\frac{\textstyle \lambda}{\textstyle 2\,\mu\,(\lambda+\mu)}\sigma_{xx}\,\sigma_{zz}\right\}\,dx\,dz.
\end{array}
\label{NRJ}
\end{equation}
At each time step, the integral in (\ref{NRJ}) is estimated by a basic trapezoidal rule at the grid nodes inside $\Omega$. Figure \ref{FigTest4}-(b) shows the time history of this mechanical energy so-obtained. It slightly decreases, due to the numerical diffusion of the scheme, which confirms that the method is stable.

\section{Conclusion}\label{SecConclu}

Here we have presented a method of incorporating free boundaries into time-domain single-grid finite-difference schemes for elastic wave simulations. This method is based on fictitious values of the solution in the vacuum, which are used by the numerical integration scheme near boundaries. These high-order fictitious values accurately describe both the boundary conditions and the geometrical features of the boundaries. The method is robust, involving negligible extra computational costs. 

Unlike the vacuum method, the quality of the numerical solution thus obtained is almost independent of the angle between the free boundaries and the Cartesian meshing. Since the free boundaries do not introduce any additional artefacts, one can use the same discretization as in homogeneous media. Typically, when a fourth-order ADER scheme is used on a propagation distance of 50 minimal wavelengths, 10 grid nodes per minimal S-wavelength yield to a very good level of accuracy. With 5 grid nodes per minimal S-wavelength, the solution is less accurate but still acceptable. 

For the sake of simplicity, we have dealt here with academic cases, considering two-dimensional geometries, constant physical parameters, and simple elastic media. Let us examine briefly the generalization of our approach to more realistic configurations:
\begin{enumerate}
\item Extending the method to 3-D topographies a priori does not require new tools. The main challenge will concern the computational efficiency of parallelization. A key point is that the determination of each fictitious value is local, using numerical values only at neighboring grid nodes. Particular care will however be required for fictitious values near frontiers between computational subdomains, in order to minimize the exchanges of data.
\item Near free boundaries, the domains of propagation are usually smoothly heterogeneous. To generalize our method to continuously variable media, the main novelty expected concerns the high-order boundary conditions detailed in section \ref{SubSecBC}. With variable matrices $\mitbf{A}$ and $\mitbf{B}$ indeed and $k\geq 2$, the procedure (\ref{DTauDt}) will involve the following quantities, to be estimated numerically:
$$
\frac{\textstyle \partial^{k-1}}{\textstyle \partial\,x^{k-1-\alpha}\,\partial\,z^\alpha}\,\mitbf{A},\quad \frac{\textstyle \partial^{k-1}}{\textstyle \partial\,x^{k-1-\alpha}\,\partial\,z^\alpha}\,\mitbf{B},\qquad \alpha=0,...,k-1.
$$
\item Realistic modeling of wave propagation requires to incorporate attenuation. The only rheological viscoelastic models able to approximate constant quality factor over a frequency range are the generalized Maxwell body \citep{EMMERICH84} and the generalized Zener body \citep{CARCIONE01}. These two equivalent models \citep{MOCZO05} yield to additional unknowns called \textit{memory variables}. In the time domain, the whole set of unknowns satisfies a linear hyperbolic system with source term
\begin{equation}
\frac{\textstyle \partial}{\textstyle \partial\,t}\,\mitbf{U}=\mitbf{A}\,\frac{\textstyle \partial}{\textstyle \partial\,x}\,\mitbf{U}+\mitbf{B}\,\frac{\textstyle \partial}{\textstyle \partial\,z}\,\mitbf{U}-\mitbf{S}\,\mitbf{U},
\label{LCatten}
\end{equation}
where $\mitbf{S}$ is a definite positive matrix. Compared with the elastic case (\ref{LC}) examined in the present paper, the main difference expected concerns the time differentiation of the boundary condition (\ref{L0U0}). Indeed, equation (\ref{L0t}) has to be modified accordingly to (\ref{LCatten}). Similar modifications are also foreseen in the case of poroelasticity in the low-frequency range \citep{DAI95}, where the evolution equations can be put in the form (\ref{LCatten}).
 \end{enumerate}

\label{lastpage}

\begin{acknowledgments}
The authors thank the reviewers P. Moczo and I. Oprsal for their instructive comments and bibliographic insights.
\end{acknowledgments}

\def\newblock{\hskip .11em plus .33em minus .07em}
\bibliographystyle{gji}
\bibliography{biblio}

\end{document}